\documentclass[useAMS,usenatbib]{mn2e}
\usepackage{graphicx}
\usepackage{natbib}
\usepackage{amsmath}
\usepackage{times}
\usepackage{float}
\usepackage{epsfig}

\newcommand{\be}{\begin{equation}}
\newcommand{\ee}{\end{equation}}

\title{The  TRAPPIST-1 system:
Orbital evolution, tidal dissipation, formation and habitability} 

\author{John C. B. Papaloizou}
\author[J. C. B.  Papaloizou et al.]{ J.C.B.Papaloizou$^{1}$\thanks{Email:jcbp2@dampt.cam.ac.uk},
Ewa Szuszkiewicz$^{2}$\thanks{Email:szusz@feynman.fiz.univ.szczecin.pl},
 Caroline Terquem$^{3}$\thanks{Email:caroline.terquem@physics.ox.ac.uk}\\
$^{1}$DAMTP, University of Cambridge, Wilberforce Road, Cambridge, CB3 0WA, U.K.\\
$^{2}$Institute
of Physics and CASA$^*$, Faculty of Mathematics and Physics, University of Szczecin, ul. Wielkopolska 15, 70-451 Szczecin, Poland.\\
$^{3}$Physics Department, University of Oxford, Keble Road, Oxford OX1 3RH, U.K.}
\begin{document}

\maketitle

\begin{abstract}
We study the dynamical evolution of the TRAPPIST-1 system under the influence of orbital circularization
through tidal interaction with the central star. We find that systems with  parameters  close to the observed one
evolve into a state where consecutive planets are linked by first order resonances and consecutive triples,     
apart  from planets c, d and e,  by connected three body Laplace resonances. The system  expands with period ratios
increasing  and mean eccentricities decreasing with time.  
This evolution is largely driven by tides acting on the innermost planets
which  then  influence the outer ones. In order  that deviations from commensurability become significant   
 only on $Gy$ time scales or longer, we require that the tidal parameter  associated with the planets has to be such that
  $Q'  > \sim 10^{2 -3}.$ 
 At the same time, if we start with
two subsystems, with the inner three planets comprising the inner one, 
 $Q'$  associated with the planets  has to be on the order (and not significantly exceeding)
 $10^{2-3}$  for the two subsystems to interact and end up in the observed
configuration.
This scenario is also supported by modelling of the evolution through disk  migration  which indicates  that the whole system cannot have migrated inwards
together. Also in order to avoid                                                                    
large departures from commensurabilities, the system cannot have stalled at a disk inner edge for significant time periods. 
We discuss the habitability consequences of the tidal dissipation implied by our modelling, concluding that planets  
d, e and f are potentially in  habitable zones.

\end{abstract}


\section{Introduction}\label{sec1}
The TRAPPIST-1 system is the first transiting planet system found orbiting an ultra-cool
M$8V$
dwarf star. Transits of  the planets  TRAPPIST-1 b, c and d  with radii comparable to the earth  were reported by Gillon et al. (2016).
 Subsequently the number of transiting planets was increased to seven by Gillon et al. (2017).
 They noted that the inner six planets were  in near commensurable orbits and thus form a resonant chain.
 Luger et al. (2017) noted the possibility of three body resonances amongst sets of consecutive triples
 and  postulated the  existence of a three body resonance between planets f, g and h, 
 which led  to the specification of a 3:2 resonance between planets g and h,  which enabled a  firm detection of the period
 of planet h. They also indicated the existence of a three body Laplace resonance between planets b, c and d,
 which can be thought of resulting from 3:2  commensurabilities  between the pair b and c and  between  the pair c and d,  even
  though departures from these first order  commensurabilities  are large.
  Allowing for   this situation,  all consecutive pairs are in or affected by  first order commensurabilities (see table \ref{table1} below).
  
  The configuration of TRAPPIST-1 is indicative that it was formed by inward migration through a protoplanetary disc
  as resonant chains form naturally in that  scenario (eg. Cresswell \& Nelson  2006). In addition they are close enough to the central star
   for it to be possible that tidal interaction has affected the dynamics and led to evolution of successive period ratios (eg. Papaloizou 2011).
   As noted by Gillon et al. (2017) conditions are such that planets  d - g may have liquid water on their surfaces and therefore
   possibly be habitable.  
 All of these features make the TRAPPIST-1 system one of topical interest.
 
 In this paper we study the dynamics of the TRAPPIST-1
system taking into account tidal interaction with the central star
that leads to orbital circularization.  This allows us to obtain
constraints on the strength of this interaction that come about in
order to avoid significant departures from the present configuration
over its estimated lifetime.
We also consider the consequences of the  tidal dissipation  that could  occur when these constraints are adhered to
for the habitability of the planets in the system.

As the system  is likely to have reached its present configuration as a result  of migration  from larger distances in a protoplanetary 
disc, we construct models for this process that adopt the expected scaling for migration rates through the disc.
In particular we consider the issue as to whether the system can have migrated as a unit or separated into two detached 
subsystems.  In the  latter  case  we perform simulations to determine under what conditions evolution under tidal interaction
with the central star can have restored the system to its current form.   

The plan of this paper is as follows.
In Section \ref{sec2} we  describe the physical model and basic equations we use
to describe the planetary system moving under its mutual gravitational interactions
together with forces due to tidal interaction with the central star and interaction with a protoplanetary disc if present.  
In Section \ref{SLap} we give an analytical description of a system of N planets,
in which consecutive groups of three planets are linked by three body
Laplace resonances, which expands under the action of tidal
circularization
which occurs while the system conserves angular momentum. This expansion
causes a secular increase of  the  period ratios seen in simulations.
In Section \ref{twosubsystems} the treatment is extended to consider two connected subsystems.

We  go on to describe our numerical simulations of two model systems, A,  and B,  with
orbital  parameters close to those
of the present TRAPPIST-1 system, but different masses in Section \ref{Numsimsec}.
We give results in Section \ref{Numressec} that show  that the evolution  in general  sets up a system
of three body Laplace resonances  with  related pairs of first order resonances 
 connecting all  consecutive sets of three planets,  apart from
planets c, d and e for different values of  their associated tidal parameter $Q'.$ 
We go on to consider
simulations  for which either the inner three planets  or the inner two  were split off
from the rest and moved to smaller radii, so forming a separate subsystem 
in section \ref{Intsub}. These indicate that  this subsystem can merge with that formed by the outer planets 
to form a system like TRAPPIST-1.

In order to investigate the origin of the system we perform simulations of
the planets migrating in a protoplanetary disc in Section \ref{disceffect}.
We  also consider the maintenance of commensurabilities, the evolution of period ratios  and the effects of disc dispersal
on  resonant chains.
 We discuss the possible effects of the tidal dissipation inferred from our simulations
on habitability in Section \ref{Habitability}.
 Finally we summarise and discuss our results  in Section \ref{sec:discussion}.

\section{Model and basic equations}\label{sec2}
We consider a system of $N$ planets moving under their mutual gravitational attraction and that due to the central star.
The equations of motion are:
\begin{equation}
{d^2 {\bf r}_i\over dt^2} = -{GM{\bf r}_i\over |{\bf r}_i|^3}
-\sum_{j=1\ne i}^N {Gm_j  \left({\bf r}_i-{\bf r}_j \right) \over |{\bf
    r}_i-{\bf r}_j |^3} -{\bf \Gamma}  +{\bf \Gamma}_{m,i}   +{\bf \Gamma}_{e,i} +{\bf \Gamma}_{r,i} \; ,
\label{emot}
\end{equation}

\noindent  where $M$, $m_i,$ $m_j,$ ${\bf r}_i$ and ${\bf r}_j$ denote the mass of
the central star, the mass of planet~$i,$ the mass of planet~$j,$  the position vector of planet
$i$ and the position vector of planet
$j,$ respectively. 
The enumeration is such  that $i=1$ corresponds to the innermost planet and $i=N$ to the outermost planet.
 The acceleration of the coordinate system based on
the central star (indirect term) is:
\begin{equation}
{\bf \Gamma}= \sum_{j=1}^N {Gm_j{\bf r}_{j} \over |{\bf r}_{j}|^3}.
\label{indt}
\end{equation}
Migration torques can be  included through,
${\bf \Gamma}_{m,i},$
which takes the form (see eg. Terquem \& Papaloizou 2007)
\begin{align}
&{\bf \Gamma}_{m,i} =
 \frac{ 
{\bf r}_i\times \left({\bf r}_i\times d {\bf r}_i/dt \right) }{3 |{\bf r}_i|^2 t_{mig,i}}
\label{Mui}
\end{align}
where $t_{mig,i}$
is defined to be the inward  migration time for planet $i,$
being the characteristic time scale on which the mean motion increases.
\noindent Tidal interaction with  the central star
 is dealt with through the addition of a  frictional  damping force taking the form (see eg.
Papaloizou~\& Terquem~2010)
\begin{equation}
{\bf \Gamma}_{e,i} =
 - \frac{2}{|{\bf r}_i|^2 t_{e,i}} \left( \frac{d {\bf r}_i}{dt} \cdot
{\bf r}_i \right) {\bf r}_i 
\label{Gammai}
\end{equation}
\noindent where $t_{e,i}$  is  the time scale
over which  the eccentricity  of an isolated planet damps.
Thus it is  the orbital  circularization time (see below).
Note that  in this 
formulation, eccentricity damping acts through radial velocity damping,
which is associated with  energy loss at constant angular momentum.  Thus
 it causes  both the semi--major axis and the eccentricity to decrease.
  We write
 \begin{equation}
 \frac{1}{t_{e,i}} = \frac{1}{t_{e,i}^d} + \frac{1}{t_{e,i}^s} ,
 \end{equation}
 \noindent where $t_{e,i}^d$ and $t_{e,i}^s$ are the damping timescales   due to
the protoplanetary disc and tidal circularization due to the central star respectively.
\noindent  Relativistic
effects  may be included through ${\bf \Gamma}_{r,i}$ (see Papaloizou \&
Terquem 2001).

\subsection{Time scale for orbital migration}\label{TypeImigration}\label{sec3}
It is likely that systems of close orbiting planets  were not formed in their present locations
but were formed further out and then migrated inwards while the protoplanetary disc was still present 
(see eg. Papaloizou \& Terquem  2006;  Baruteau et al.  2014  for reviews).
In particular   convergent  disc  migration readily produces   multiple systems
in resonant chains of the type observed  in TRAPPIST-1 (eg. Cresswell \& Nelson 2006;  Papaloizou \& Terquem 2010).
However,  if they form   relatively close to resonance  with migration operating only for a limited time,   the system as a whole may undergo   little net radial migration.
But note that a rapid  removal of the disc, due to  for example photoevaporation, may cause disruption of the resonances (eg. Terquem 2017). 

In this paper we assume that the system was formed  near to its current state and examine the role of tides due to the central star
in cementing the resonances and  causing subsequent evolution. This leads to constraints on the strength of the tides.
We also consider  properties of the protoplanetary disc,  and the time scale for its dispersal, that  ensure the system is not immediately disrupted.

\subsection {Time scale for orbital circularization due to tides from the central star}\label{sec4}
The circularization timescale due to tidal interaction with the star
was obtained from   Goldreich~\& Soter (1966) in the form
\begin{align}
&t_{e,i}^s =763000Q'\left(\frac{M_{\oplus}}{m_i}\right)^{2/3}\hspace{-1mm}\left(\frac{M_{\odot}}{M}\right)^{1.5}
\left(\frac{{\overline \rho}_i}{{\overline{\rho}}_{\oplus}}\right)^{5/3}\left(\frac{20a_{i}}{1AU}\right)^{6.5} y.\nonumber\\
&\equiv 763000Q'\left(\frac{m_i}{M_{\oplus}}\right)\hspace{-1mm}\left(\frac{M_{\odot}}{M}\right)^{1.5}
\left(\frac{R_{\oplus}}{R_{p,i}}\right)^{5}
\left(\frac{20a_{i}}{1AU}\right)^{6.5} y.
\label{teccs}
\end{align}

\noindent where
 $a_i$ is the 
   the semi--major axis of planet~$i,$  $R_{p,i}$  is its radius and 
 ${\overline{\rho}}_i$ is its  mean density.  The quantity
 $Q'= 3Q/(2k_2),$ where $Q$ is the tidal dissipation function and $k_2$
is the Love number.  
 For solar system planets in the terrestrial mass
range, Goldreich \& Soter (1966) give estimates for $Q$ in the range
10--500 and $k_2 \sim 0.3$, which correspond to $Q'$ in the range
50--2500. 
 However,  
 the value of $Q$ is expected
to be a function of tidal forcing frequency and temperature that could 
attain a minimum value $\sim 1$ at the solidus temperature
(see Ojakangas  \& Stevenson  1986 and  references therein). 
Although   $Q'$ is  very uncertain for extrasolar planets, we should note  that
$Q$ may be of order unity under early post formation conditions if they result in a planet being  near to  the solidus temperature. 
We remark that when circularization operates,  the rate of energy dissipation produced in planet $i,$
with orbital energy $E_i,$
is given by (see Papaloizou \& Terquem 2010):
\begin{align}
\frac{dE_i}{dt} = \frac {2e_i^2 E_i}{ t_{e,i}^s }\label{Edissip}. 
\end{align}


For  the low mass planets considered here,  we  neglect tides induced on the central star as these cannot transfer
a significant amount of angular momentum (see eg. Barnes et al. 2009a; Papaloizou \& Terquem 2010).
 Accordingly  the orbital angular momentum 
and inclination  are not changed by the tidal interaction.   
%
%
%
%

\begin{figure*}
\begin{center}
\vspace{0cm}
\hspace{-0cm}\includegraphics[width=21cm]{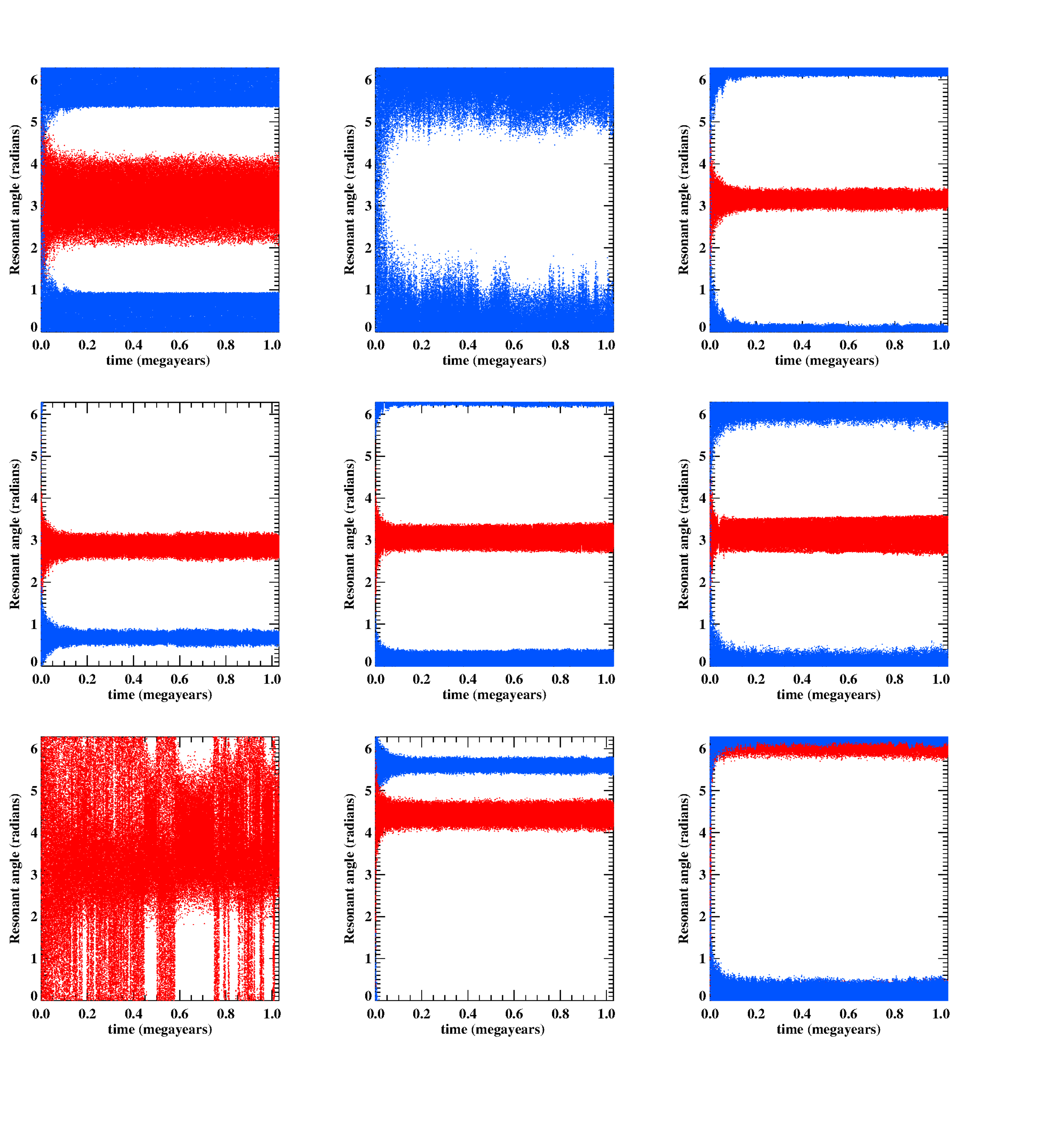} 
\end{center}
\vspace{-2cm}
 \caption{
The evolution of the seven planet system with $Q'=Q_0'=0.122127$ but tidal dissipation applied only to the innermost two planets
 and case A masses, resonant angles.
The uppermost left panel shows the resonant angles $3\lambda_2-2\lambda_1-\varpi_2$ (red curve) and  $3\lambda_2-2\lambda_1-\varpi_1$ (blue curve).
 The uppermost middle panel shows the resonant angle $3\lambda_3-2\lambda_2-\varpi_2$ and
 the uppermost right panel shows the resonant angles $3\lambda_4-2\lambda_3-\varpi_4$ (red curve) and  $3\lambda_4-2\lambda_3-\varpi_3$ (blue curve).
 The left panel in the central row shows the resonant angles $3\lambda_5-2\lambda_4-\varpi_5$ (red curve) and  $3\lambda_5-2\lambda_4-\varpi_4$ (blue curve).
 The middle  panel in the central row shows the resonant angles $4\lambda_6-3\lambda_5-\varpi_6$ (red curve) and  $4\lambda_6-3\lambda_5-\varpi_5$ (blue curve).
 The right  panel in the middle row shows the resonant angles $3\lambda_7-2\lambda_6-\varpi_7$ (red curve) and  $3\lambda_7-2\lambda_6-\varpi_6$ (blue curve).
 The lowermost left  panel shows the resonant angle $5\lambda_3-3\lambda_2-2\varpi_3.$
 The lowermost middle  panel shows the resonant angles $2\lambda_6-\lambda_4-\varpi_6$ (red curve) and  $2\lambda_6-\lambda_4-\varpi_4$ (blue curve).
The lowermost right  panel shows the resonant angles $2\lambda_7-\lambda_5-\varpi_7$ (red curve) and  $2\lambda_7-\lambda_5-\varpi_5$ (blue curve).
 \label{2Qp1angles}}
\end{figure*}
\begin{figure*}
\begin{center}
\vspace{1cm}
\hspace{-5mm}\includegraphics[width=18cm]{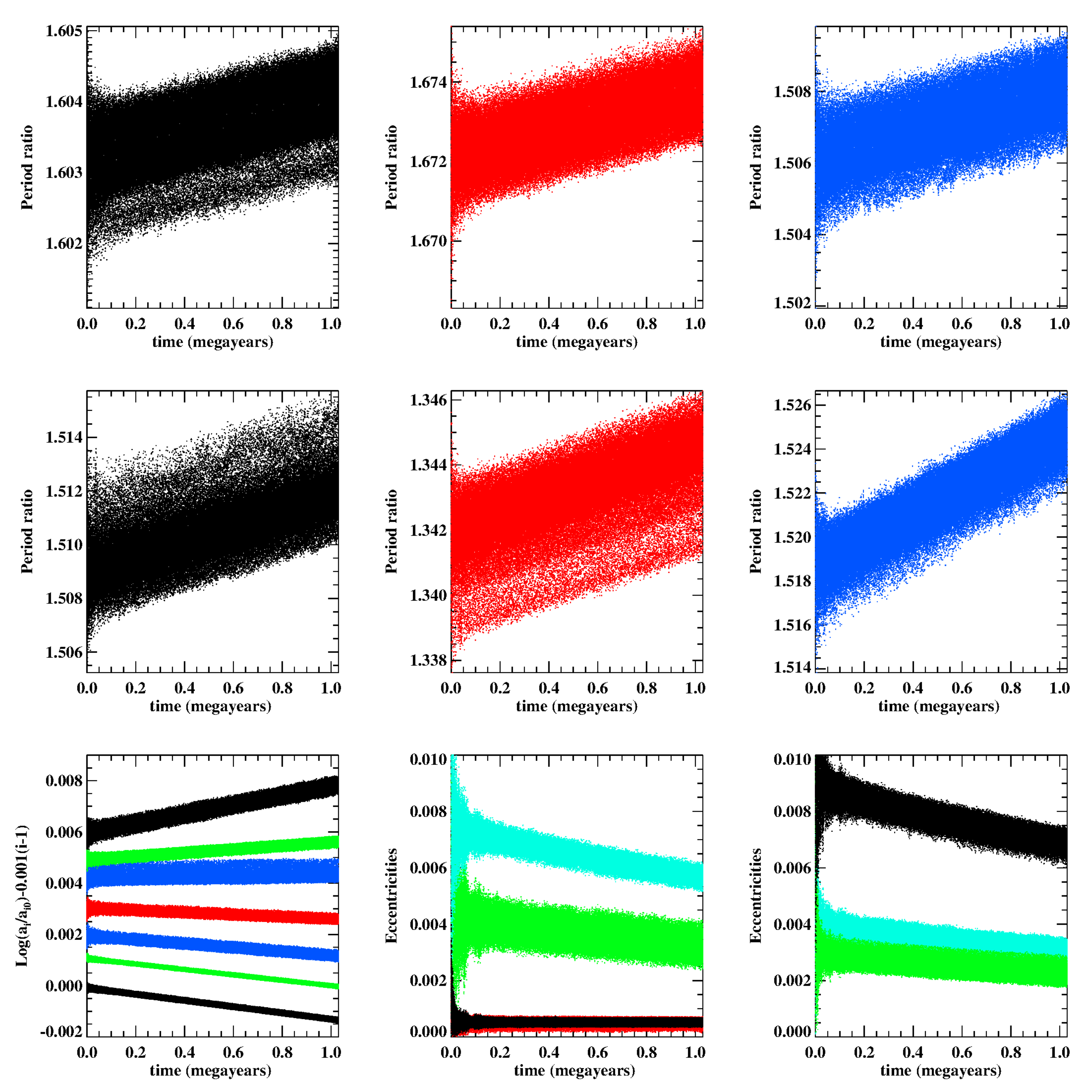} 
\end{center}
 \caption{ Results for $Q'=Q'_0$  but tidal dissipation applied only to the innermost two planets and case A masses, period ratios and eccentricities.
The uppermost left panel shows the period ratio $P_2/P_1.$
 The uppermost middle  panel shows the period ratio $P_3/P_2.$
 the uppermost right  panel shows the period ratio $P_4/P_3.$
 The left panel in the central row shows the period ratio $P_5/P_4.$
 The middle  panel in the central row shows the period ratio $P_6/P_5.$
 The right  panel in the middle row shows the period ratio $P_7/P_6.$
 The lowermost left  panel shows  $\log (a_i/a_{i0}) -0.001(i-1) $ for $i=1-7,$ where $a_{i0}$ is the initial value of the semi-major axis of planet $i.$
 Planets $1-7$ correspond to curves taking on monotonically increasing values ranging from  lowermost to uppermost.
 The lowermost middle  panel shows the eccentricities,  $e_1$ (black curve),  $e_2$ (red curve), $e_3 $(green curve) and $e_4$ (light  blue curve).
 The lowermost right  panel shows the eccentricities,  $e_5$ (black  curve), $e_6 $(light blue curve) and $e_7$ (green curve).
 \label{2Qp1periodratios}}
\end{figure*}
\begin{figure*}
\begin{center}
\vspace{0cm}
\hspace{-0cm}\includegraphics[width=21cm]{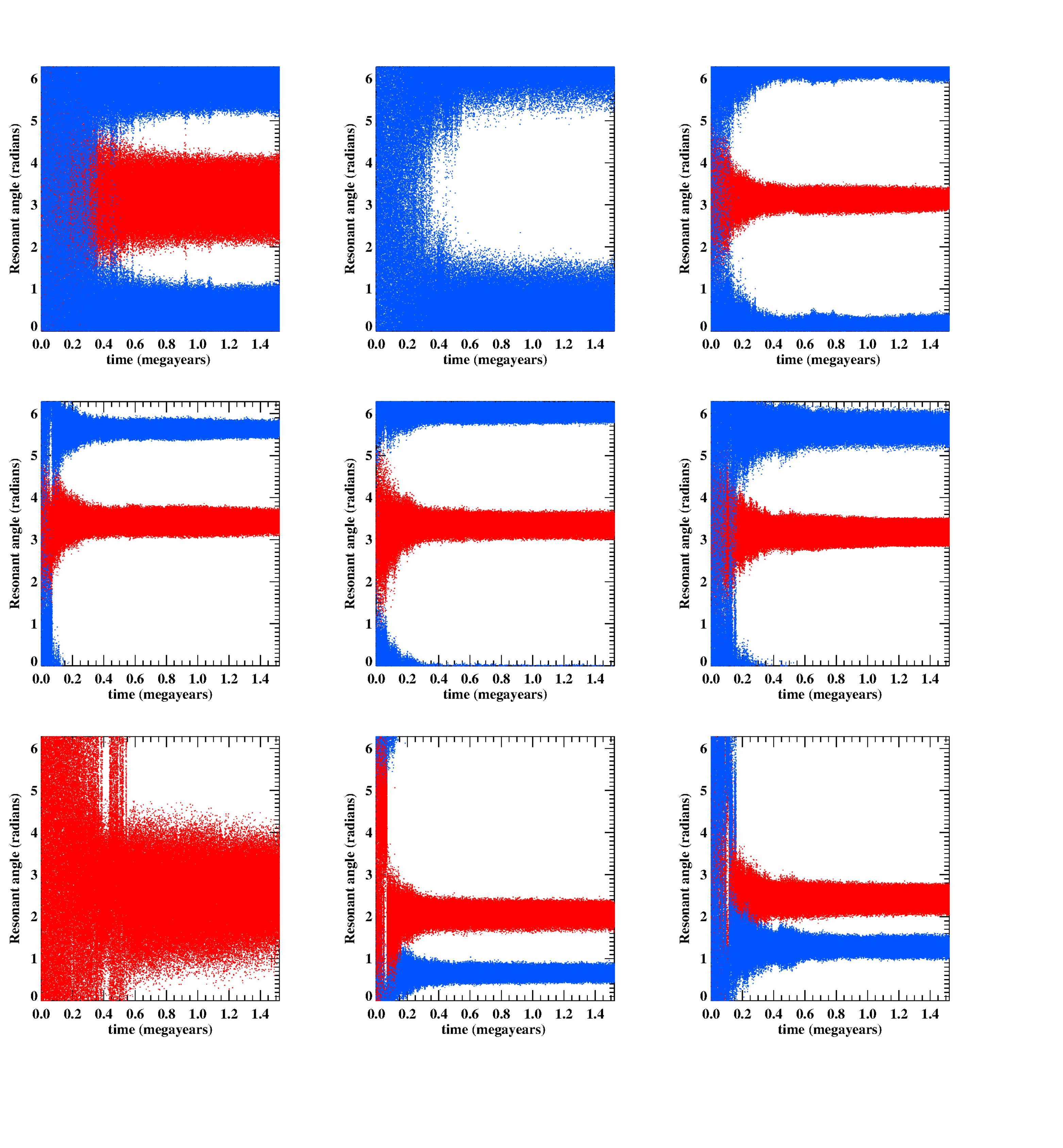} 
\end{center}
 \caption{
The evolution of the seven planet system with case A masses and  $Q'=100Q'_0$ for all planets, resonant angles.
 The contents of the panels are as in Fig. \ref{2Qp1angles}.
 \label{2Qp100angles}}
\end{figure*}
\begin{figure*}
\begin{center}
\vspace{1cm}
\hspace{-5mm}\includegraphics[width=\textwidth, height=0.8\textheight,angle=0]{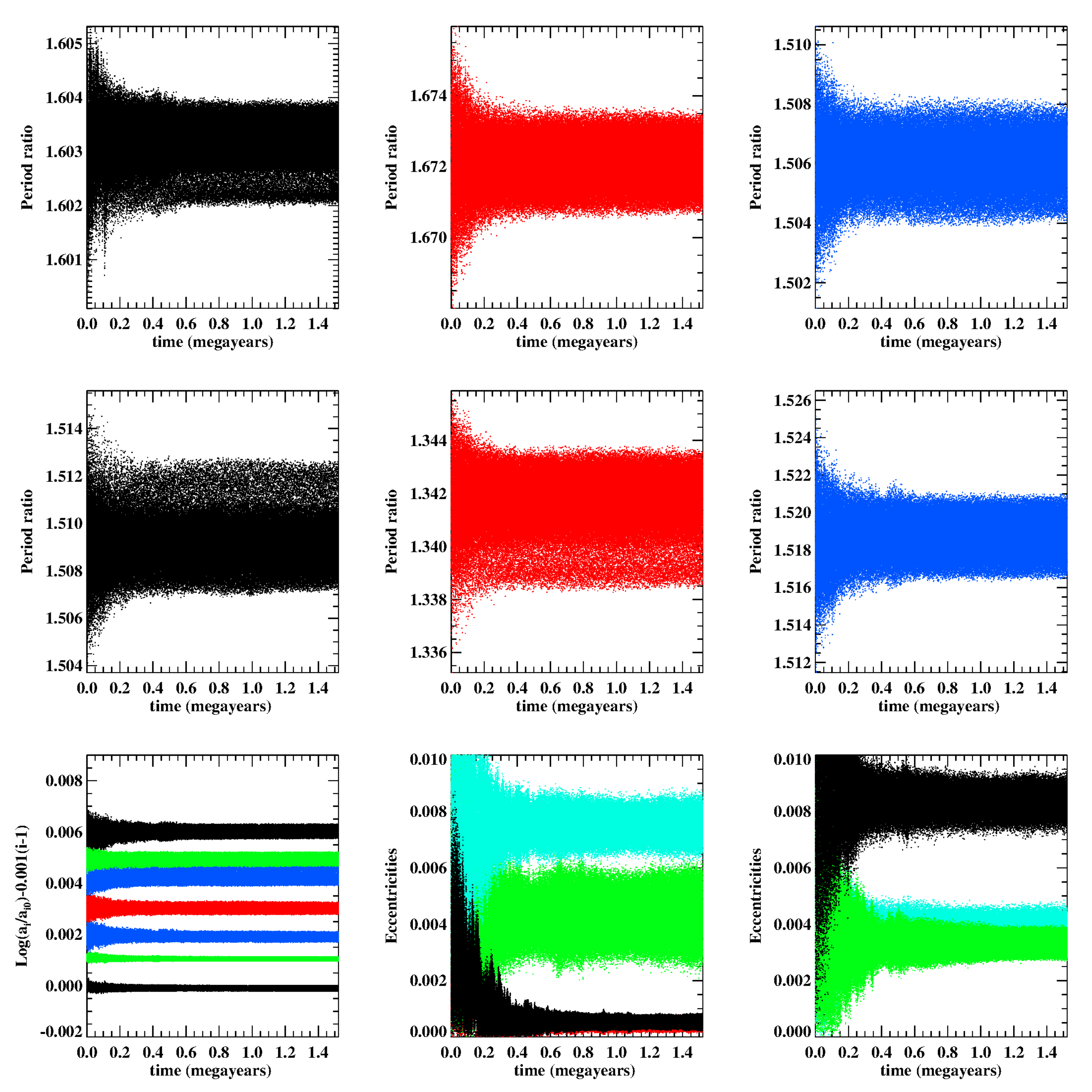}  
\end{center}
 \caption{ Results for the seven planet system with  case  A masses and $Q'=100Q'_0$ for all planets, period ratios and eccentricities.
 The contents of the panels are as in Fig. \ref{2Qp1periodratios}.
 \label{2Qp100periodratios}
}
\end{figure*}

\begin{figure*}
\begin{center}
\vspace{0cm}
\hspace{-0cm}\includegraphics[width=21cm]{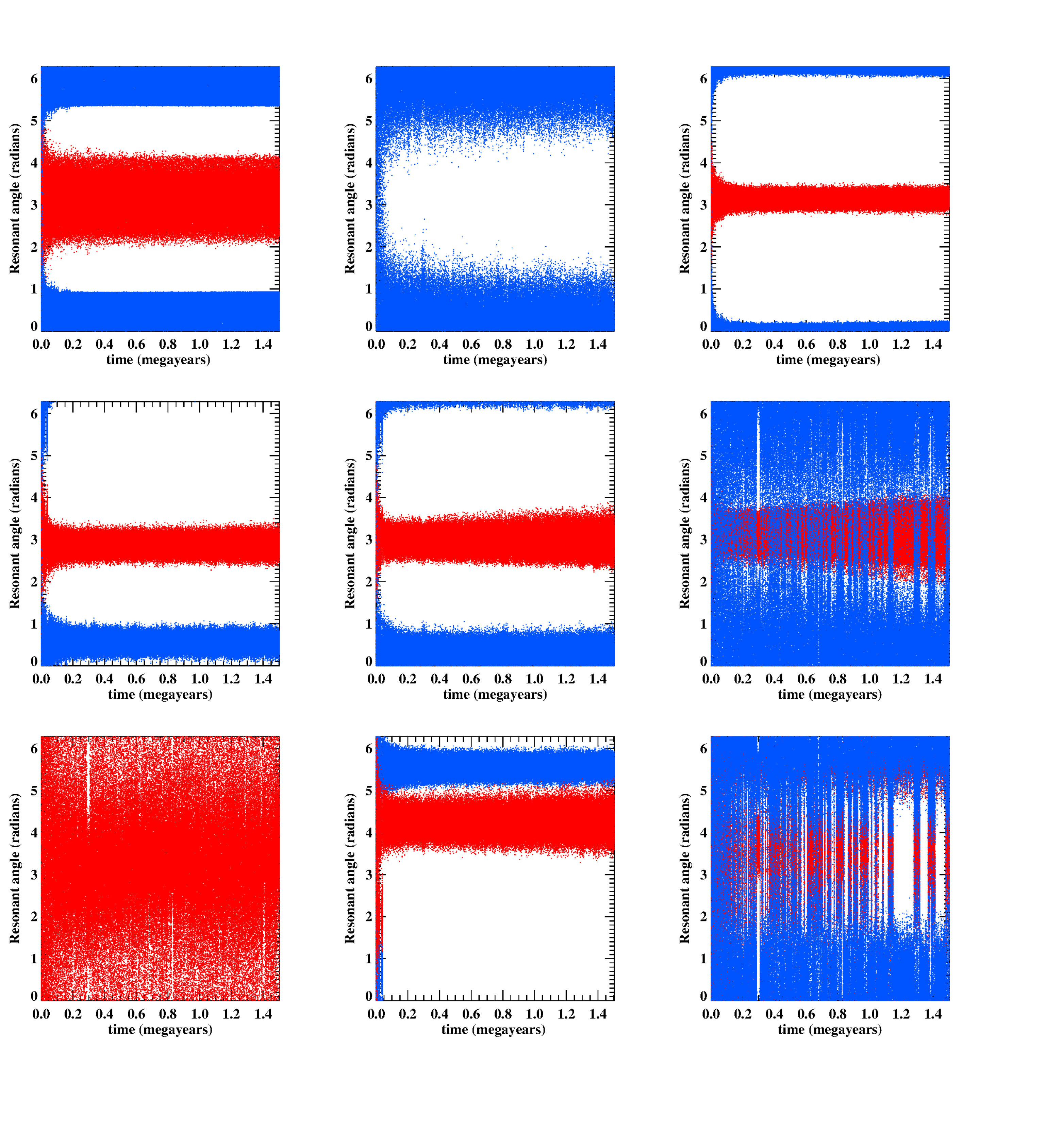} 
\end{center}
 \caption{
The evolution of the seven planet system with case B masses and  $Q'=Q'_0$   for all  planets, resonant angles.
 The contents of the panels are as in Fig. \ref{2Qp1angles}.
 \label{2Qp1newpassesangles}}
\end{figure*}

\begin{figure*}
\begin{center}
\vspace{1cm}
\hspace{-5mm}\includegraphics[width=\textwidth, height=0.8\textheight,angle=0]{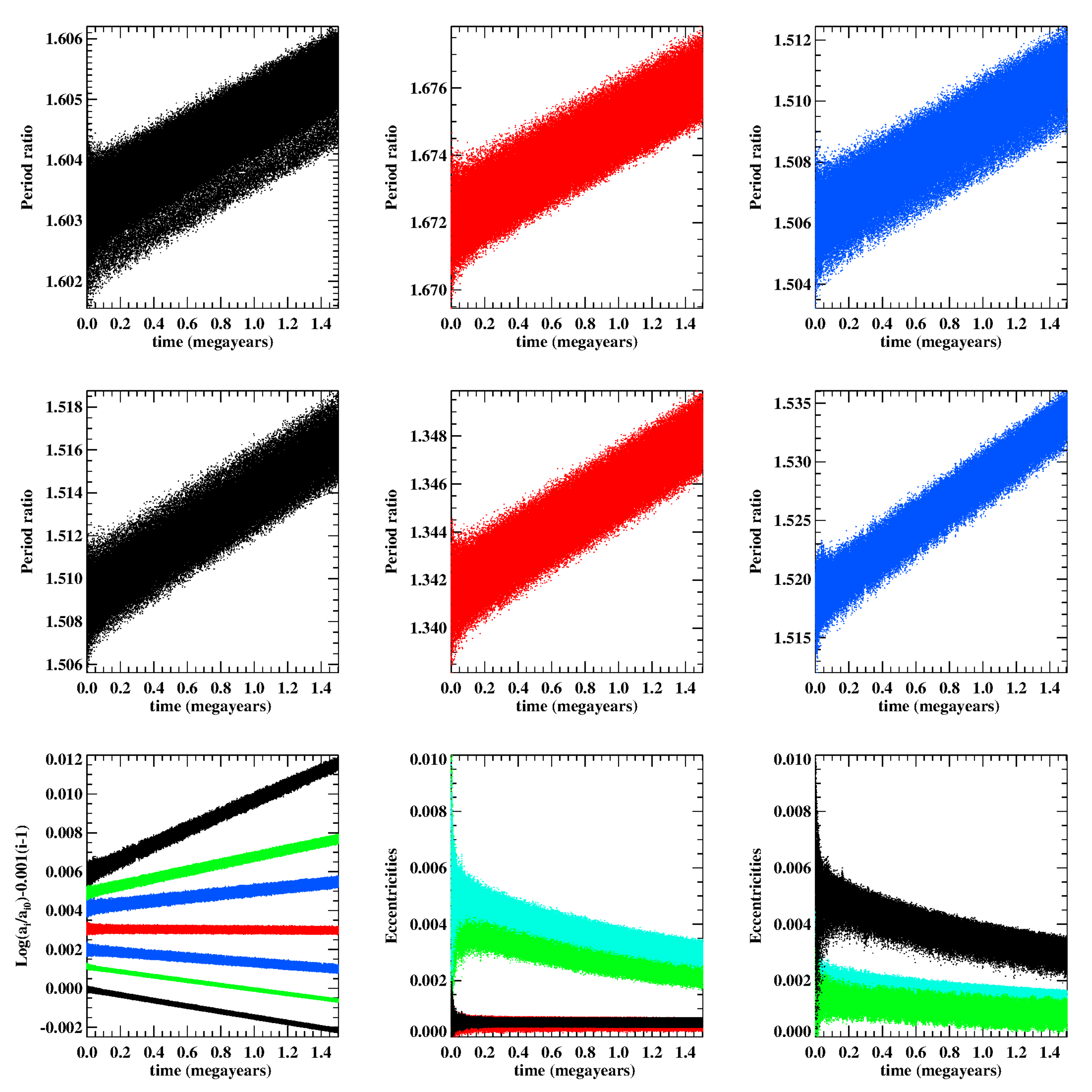}  
\end{center}
 \caption{ Results for  the seven planet system with  case  B masses and $Q'=Q'_0$ for all planets, period ratios and eccentricities.
 The contents of the panels are as in Fig. \ref{2Qp1periodratios}.
 \label{2Qp1newpassesperiodratios}
}
\end{figure*}

\begin{figure*}
\begin{center}
\vspace{0cm}
\hspace{-0cm}\includegraphics[width=21cm]{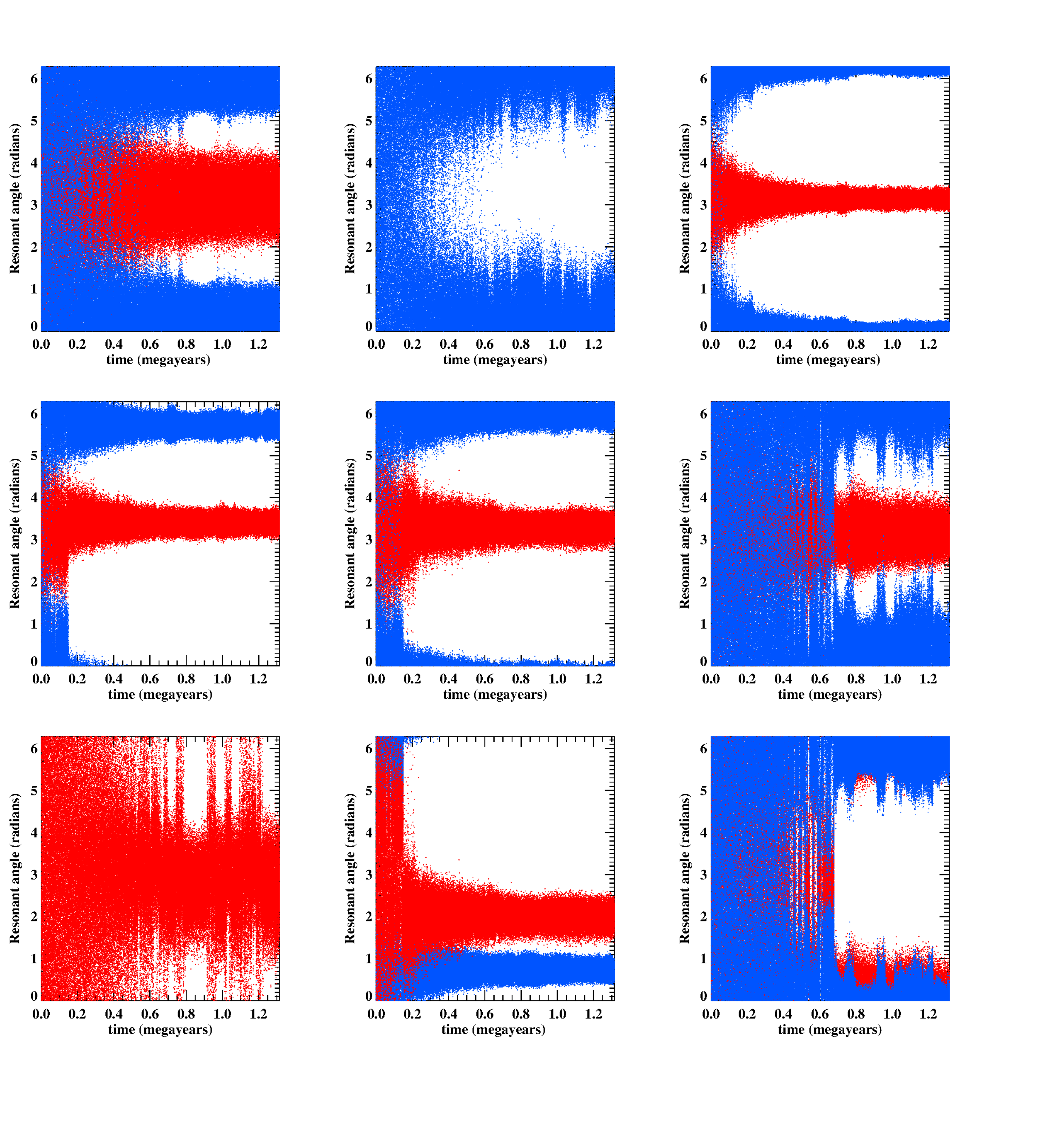} 
\end{center}
 \caption{
The evolution of the seven planet system with case B masses and  $Q'=100Q'_0$ for all planets, resonant angles.
 The contents of the panels are as in Fig. \ref{2Qp1angles}.
 \label{2Qp100newpassesangles}}
\end{figure*}
\begin{figure*}
\begin{center}
\vspace{1cm}
\hspace{-5mm}\includegraphics[width=\textwidth, height=0.8\textheight,angle=0]{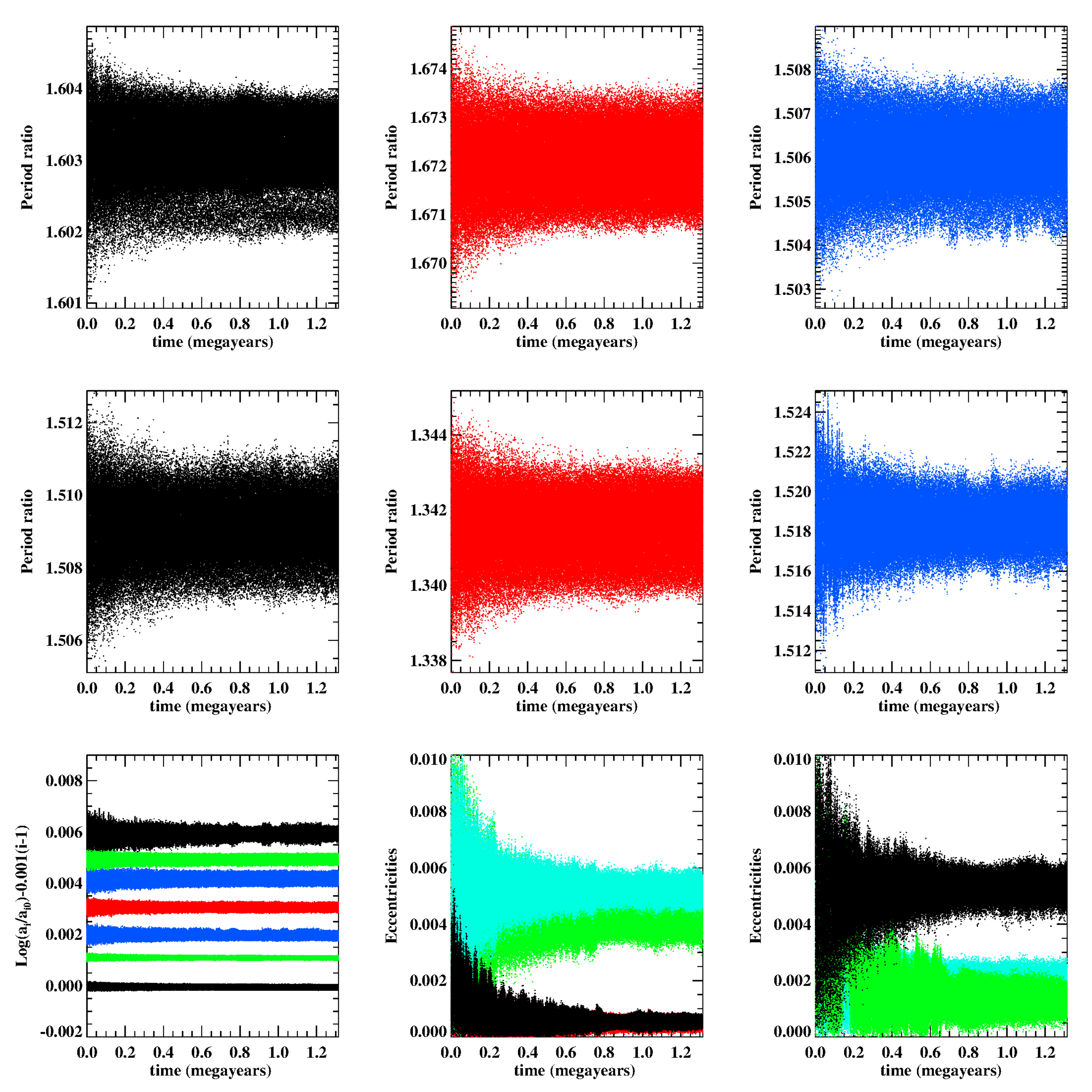}  
\end{center}
 \caption{ Results for  the seven planet system with  case  B masses and $Q'=100Q_0$ for all planets, period ratios and eccentricities.
 The contents of the panels are as in Fig. \ref{2Qp1periodratios}.
 \label{2Qp100newpassesperiodratios}
}
\end{figure*}
\begin{figure*}
\begin{center}
\vspace{0cm}
\hspace*{-0cm}\includegraphics[width=21cm]{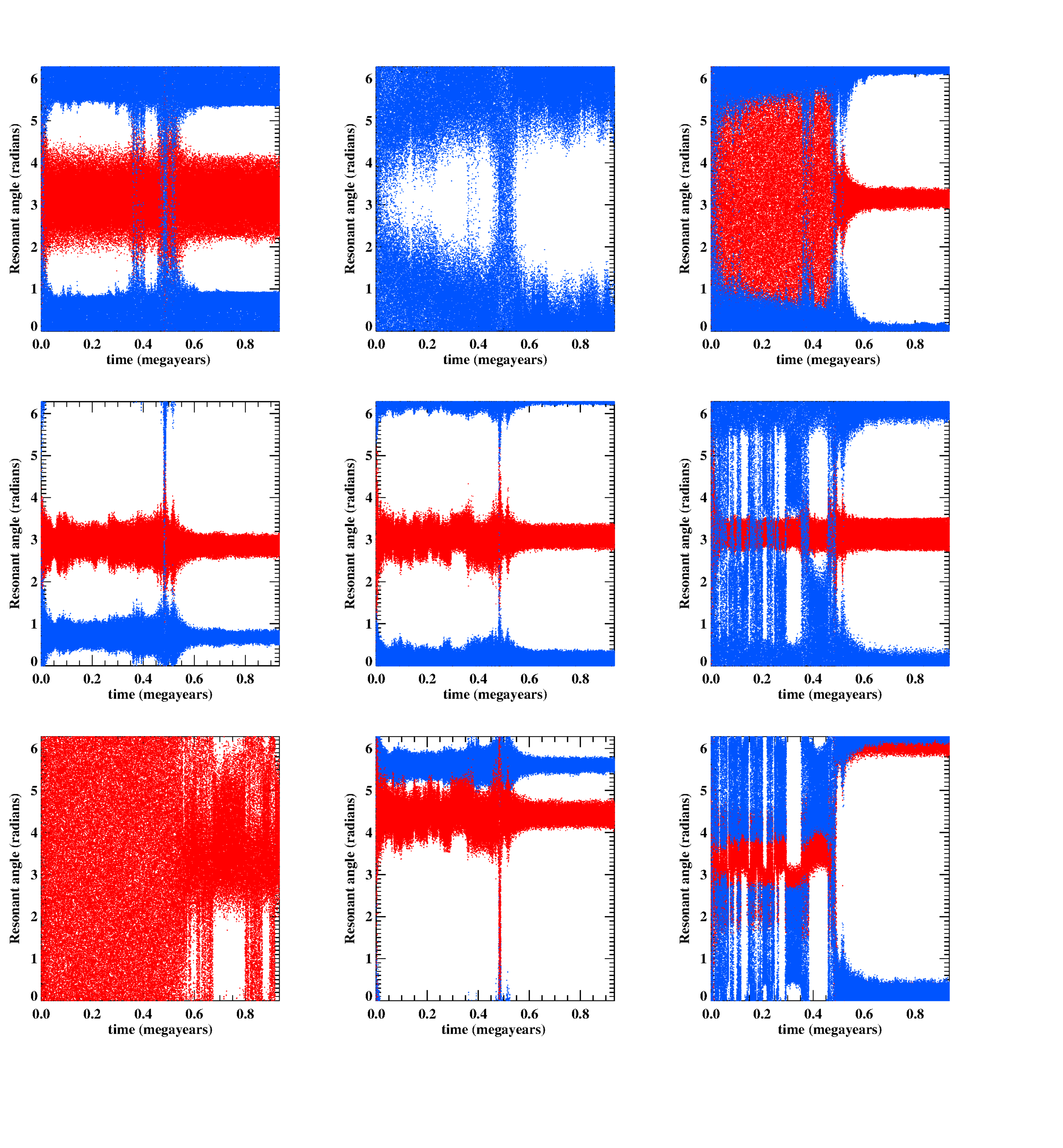} 
\end{center}
 \caption{ Results for $Q'=Q'_0$ but with inner three planets starting in a Laplace resonance with reduced semi-major axes, resonant angles.  
 The contents of the panels are as in Fig. \ref{2Qp1angles}.
 \label{2QP2angles}}
\end{figure*}
\begin{figure*}
\begin{center}
\vspace{1.5cm}
\hspace*{-5mm}\includegraphics[width=17cm, height=18cm,angle=0]{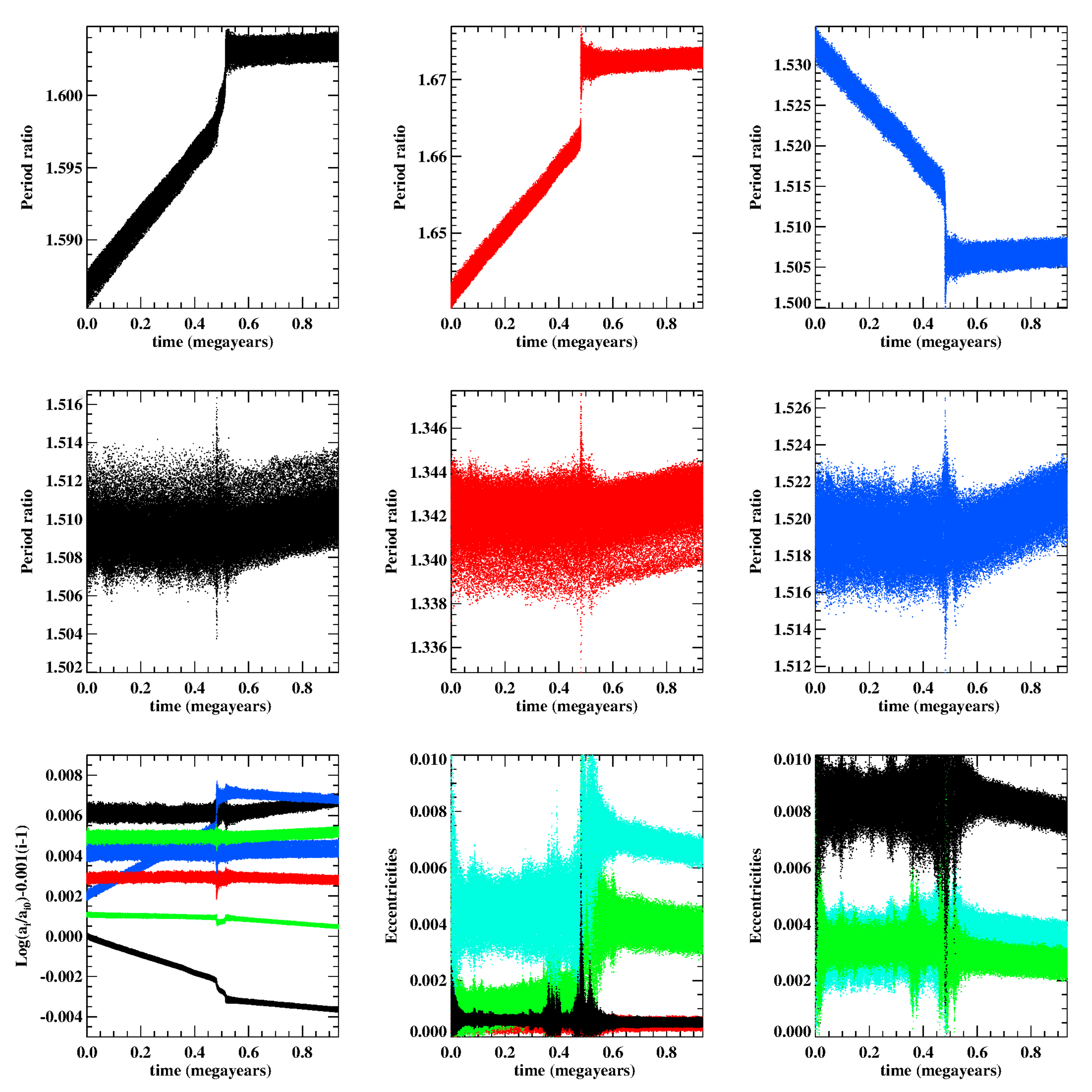} 
\end{center}
 \caption{ Results for $Q'=Q'_0$ but with inner three planets starting in a Laplace resonance with reduced semi-major axes, period ratios and eccentricities.
 The contents of the panels are as in Fig. \ref{2Qp1periodratios}.
 \label{2QP2periodratios}}
\end{figure*}
\begin{figure*}
\begin{center}
\vspace{0cm}
\hspace*{-0cm}\includegraphics[width=18cm]{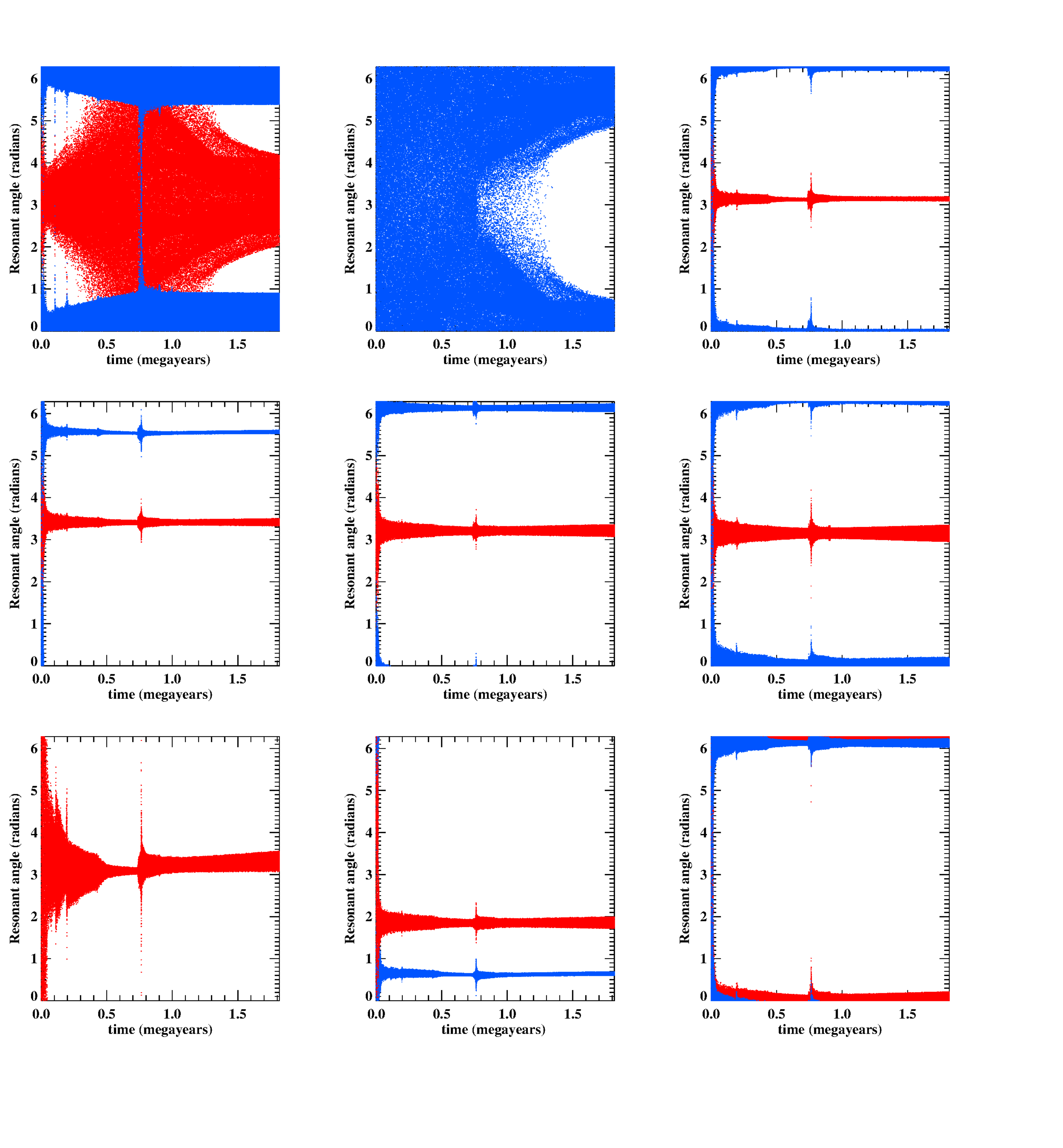} 
\end{center}
 \caption{ Results for $Q'=Q'_0$ and  masses corresponding to model A and tides acting only on the inner two planets.
Initial conditions are as for the run illustrated in Fig. \ref{2Qp1angles}
but  with inner two planets  starting in a closer  3:2 resonance with the  outermost of the pair
 at a slightly reduced semi-major axis 
,resonant angles.  
 The contents of the panels are as in Fig. \ref{2Qp1angles}. 
 \label{2Qp2anglesmig}}
\end{figure*}
\begin{figure*}
\begin{center}
\vspace{0cm}
\hspace*{-5mm}\includegraphics[width=17cm, height=18cm,angle=0]{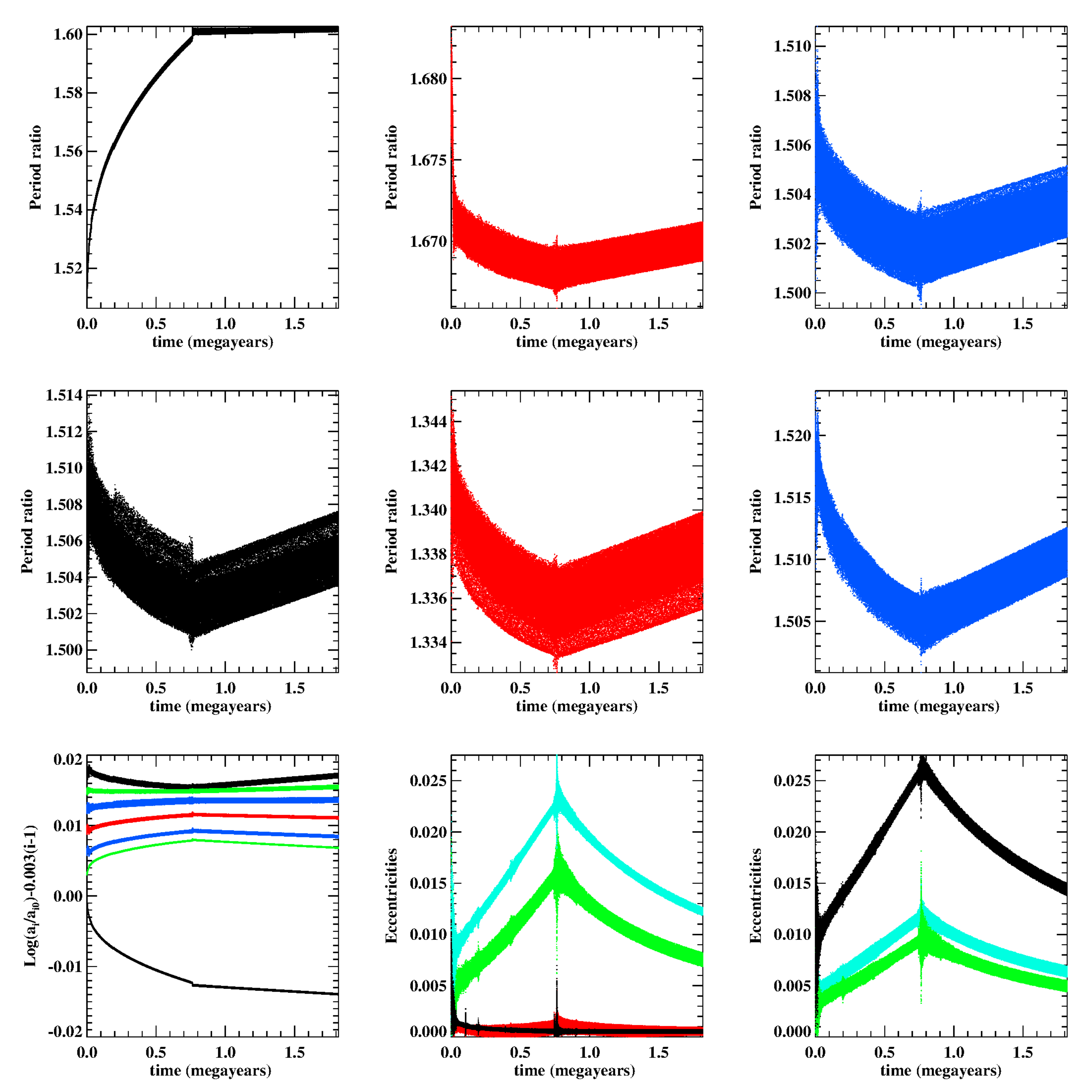} 
\end{center}
 \caption{ 
Results for $Q'=Q'_0$ and  masses corresponding to model A and tides acting only on the inner two planets.
Initial conditions are as for the run illustrated in Fig. \ref{2Qp1angles} 
but  with inner two planets  starting in a closer  3:2 resonance with the  outermost of the pair
 at a slightly reduced semi-major axis 
,period ratios and eccentricities.            
The contents of the panels are as in Fig.\ref{2Qp1periodratios}
apart from the lowermost left  panel which  shows  $\log (a_i/a_{i0}) -0.003(i-1) $ for $i=1-7,$ where $a_{i0}$ is the initial value of the semi-major axis of planet $i.$
 Planets $1-7$ correspond to curves taking on monotonically increasing values ranging from  lowermost to uppermost.
 \label{2QP2periodratiosmig}}
\end{figure*}

\begin{figure*}
\begin{center}
\includegraphics[scale=0.7]{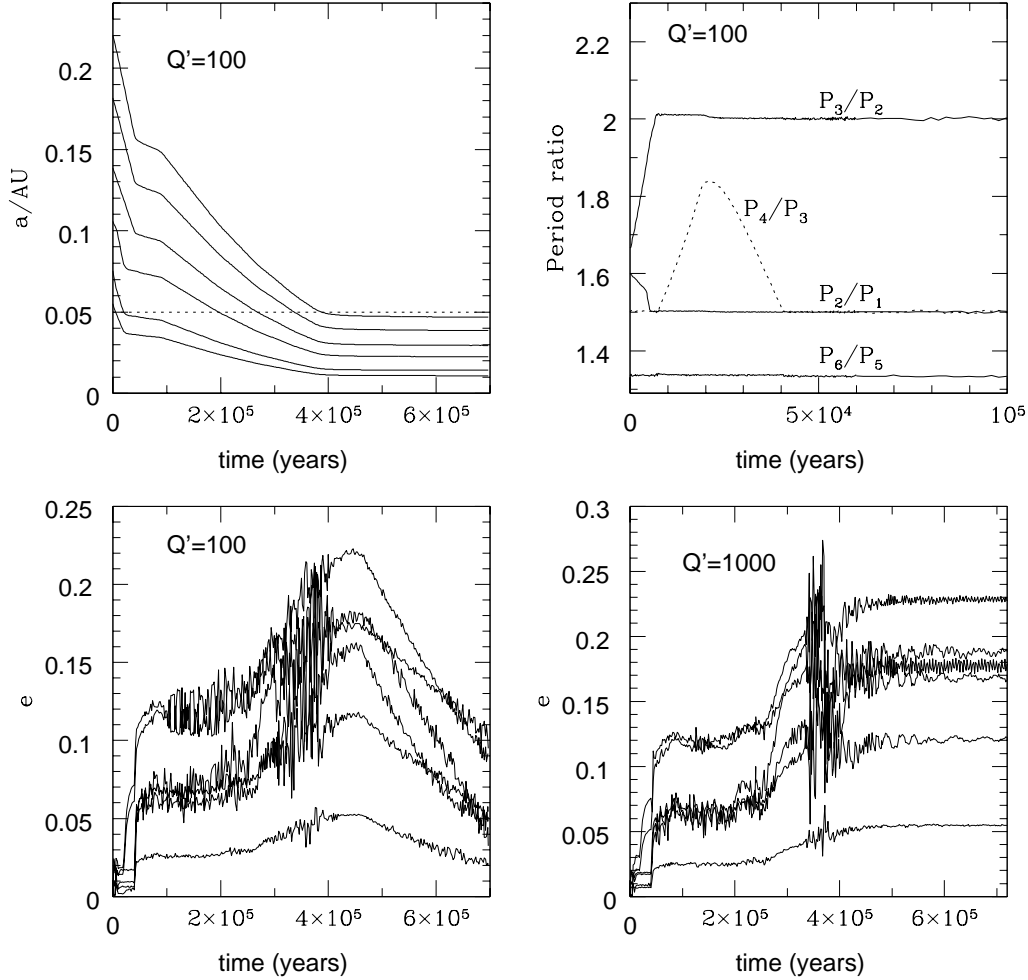}
\end{center}
\vspace{-4.cm}
\caption{Evolution of the system in case~A for
  $M_d=3.5 \times 10^{-4}$~${\rm M}_\odot$.  The upper left plot shows
  $a$ {\em versus} time for the 6 planets and Q'=100.  The upper right
  plot shows the ratio of the periods for pairs of planets {\em
    versus} time for Q'=100. Only the first $10^5$ years are shown,
  but the ratios stay almost constant after that time. The ratio
  $P_5/P_4$ is not shown but it stays constant equal to 1.5 for the
  whole duration of the simulation.  The lower plots show $e$ {\em
    versus} time for the 6 planets and Q'=100 (left panel) and Q'=1000
  (right panel). }
\label{fig:mig}
\end{figure*}

\begin{figure*}
\begin{center}
\includegraphics[scale=0.7]{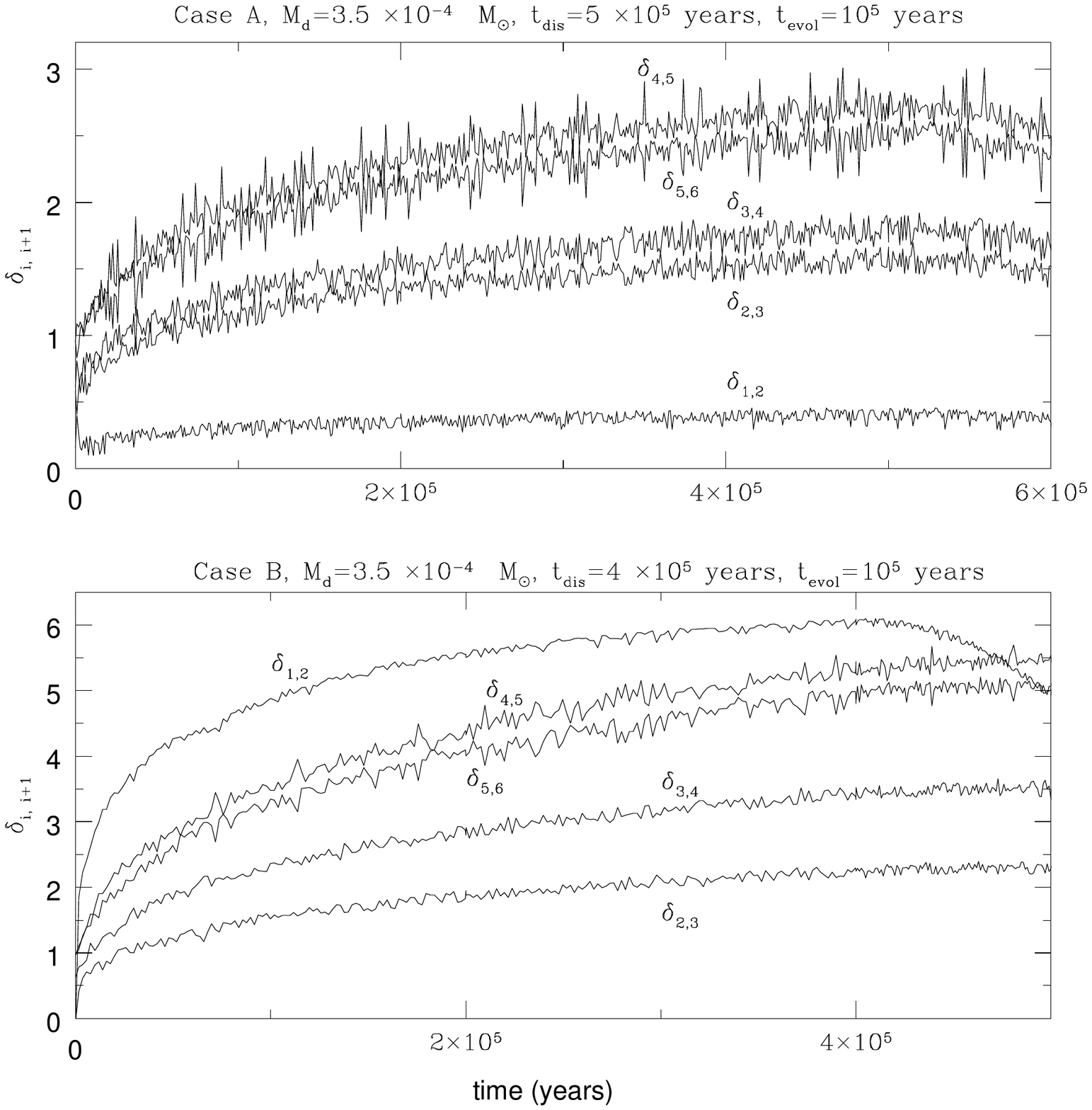}
\end{center}
\vspace{-4.cm}
\caption{Evolution of
  $\delta_{i,i+1} \equiv 100(P_{i+1}/P_{i}-r_{i,i+1})$, where
  $r_{i,i+1}$ is the initial ratio of the periods of planets $i$ and
  $i+1$.  The upper panel corresponds to case~A and
  $M_d=3.5 \times 10^{-4}$~${\rm M}_\odot$,
  $t_{\rm dis}=5 \times 10^5$~years and $t_{\rm evol}=10^5$~years. The
  lower panel corresponds to case~B and $M_d=3.5 \times
  10^{-4}$~${\rm
    M}_\odot$,
  $t_{\rm dis}=4 \times 10^5$~years and $t_{\rm evol}=10^5$~years. The
  value $\delta=1$ means a departure from exact commensurability of
  1\%. Note that the sharp increase of $\delta$ at the beginning of
  the simulations may be due to the initial set up, in which the
  resonant angles are not librating. }
\label{fig:stal}
\end{figure*}

\section{The expansion   of  multi-planet  systems  linked through a series of 
 Laplace resonances driven by  energy dissipation at fixed angular momentum}\label{SLap}

We here consider systems with consecutive planets linked by first order resonances 
and consecutive triples linked by Laplace resonances, though the latter may not apply  in detail to all triples as in the TRAPPIST-1 system.
We investigate how energy dissipation due to tidal circularization that acts on forced eccentricities but occurs while the total angular momentum
of the system is conserved, leads to a secular expansion of the system  with successive period ratios becoming increasingly disparate.
We relate the rate of total energy dissipation to the rate of expansion
when either all triples are linked by Laplace resonances, or when the system is composed of two subsystems 
each of which  are  fully linked by Laplace resonances. The evolution time scale for this process scales as $Q'.$
  
In order to proceed we  set
the  mean motion of planet $i$  to be $n_{i},$ it's orbital period to  be $P_i =2\pi/n_i, $ it's eccentricity to be  $e_i,$ the longitude of pericentre  to be  $\varpi_i,$
and its longitude to be $\lambda_i.$
The systems we consider are such that consecutive planets are in  first order resonances
for which either one or two  of the 
 associated resonant angles librate. When both resonant angles librate for every pair, consecutive triples
are linked through three body Laplace resonances  (~see eg. Papaloizou 2015, 2016~).

If  planets, $i,$ and, $i+1,$  are in the first order $p_i:(p_i-1)$ resonance, with,  $p_i,$  being an integer that may be different for different pairs of planets,
the relevant resonant angles are given by
\begin{align}
&\Phi_{i,i+1,1}  = p_{i}\lambda_{i+1}-(p_{i}-1)\lambda_i -\varpi_{i}, \hspace {3mm} {\rm  and}\nonumber\\
&\Phi_{i,i+1,2}  = p_{i}\lambda_{i+1}-(p_{i}-1)\lambda_i -\varpi_{i+i}. 
\end{align}
\begin{align}
 &\hspace{-3mm}{\rm The \hspace{1mm} angle} \hspace{1mm} \Psi_{i+2,i+1,i}= \Phi_{i+1,i+2,1} -\Phi_{i,i+1,2}\hspace{1mm}\nonumber\\
 & \hspace{-3mm} {\rm which \hspace{1mm} can \hspace{1mm} also \hspace{1mm} be\ \hspace{1mm} written}\nonumber\\
 &\Psi_{i+2,i+1,i} =  p_{i+1}\lambda_{i+2} - (p_{i}+p_{i+1}-1)\lambda_{i+1} +(p_{i}-1)\lambda_i \nonumber
 \end{align}
 \noindent  depends only on the longitudes and for it's time derivative we have
\begin{align}
&\frac{d \Psi_{i+2,i+1,i}}{dt}  = p_{i+1}n_{i+2} - (p_{i}+p_{i+1}-1)n_{i+1} +(p_{i}-1)n_i \nonumber\\
&+O\left(\frac{m_i e_i}{M}\right)
\end{align}
(see eg. Papaloizou 2015, 2016).  Thus if the  two  associated first order resonant angles   undergo small amplitude librations,  given small eccentricities
and mass ratios,  this  angle will also  undergo small amplitude librations  and  the Laplace resonance condition
\begin{equation}
p_{i+1}n_{i+2} - (p_{i}+p_{i+1}-1)n_{i+1} +(p_{i} -1)n_i =0. \label{Laplacerel}
\end {equation}
will hold. For large amplitude librations it  will hold in a  time average sense and we shall assume that the orbital elements 
we use in the simple modelling   carried out  below also correspond to appropriate time averages.

For a multi-planet system with $N > 3$ members, there can be up to $N-2$ Laplace resonance conditions 
that hold, each one connecting three consecutive planets. When all of these apply,  all but two of the semi-major axes are determined by the resonance
conditions. When this occurs we  exploit it  by introducing  the quantities 
\begin{equation}
 \Delta_{i} = n_{i}-n_{i+1}, \hspace{3mm} {\rm for}\hspace{3mm} i =1,  2, ..  N-1. 
\end{equation}
Then 
\begin{equation}
\frac{\Delta_{i+1}}{\Delta_{i}} = \frac{n_{i+1}-n_{i+2}}{n_{i}-n_{i+1}}= \frac{p_i-1}{p_{i+1}}
\end {equation}
and we have
\begin{equation}
\Theta_{i,k} =  \frac{\Delta_{i+k}}{\Delta_{i}} = \prod _{j=1}^{k}\frac{ (p_{i+j-1}-1)} {p_{i+j}}  \hspace{3mm} {\rm for} \hspace{3mm} k=1,2... N-i-1.
\label{nomenc}
\end {equation}
together with the specification $\Theta_{i,0} =1$.

Using (\ref{nomenc}) for $i=1,$ all of the mean motions  in the system  can be  expressed in terms of   $n_1$ and  $\Delta_1.$ 
Thus
\begin{equation}
n_{i}-n_{i+k+1}= \Delta_{i} \sum_{j=0}^{k}\Theta_{i,j}\label{nomenca}
\end{equation}

Differentiating (\ref{nomenca}) with respect to time we obtain
\begin{align}
& \frac{d E_{i+k+1}}{dt }  - \left(\frac{m_{i+k+1}}{m_{i}} \right) 
\left(\frac{ a_{i+k+1}} {a_{i}}\right)^{1/2}\frac{d E_{i}}{dt} =\nonumber\\
& \frac{m_{i+k+1}}{3}\sqrt{GMa_{i+k+1}}
\frac{d\Delta_{i }}{dt}\sum_{j=0}^k \Theta_{i,j}\nonumber  \\
& \hspace{-0cm} {\rm for}\hspace{3mm}  k = 0,1, ... N - i - 1.\label{nomenc2}
\end{align}
Writing this in terms of  the angular  momenta under the assumption that eccentricities are vanishingly small,  we obtain
\begin{align}
& \frac{d J_{i+k+1}}{dt }  - \left(\frac{m_{i+k+1}}{m_{i}} \right) \left(\frac{ a_{i+k+1}} {a_{i}}\right)^{2}\frac{d J_{i}}{dt} = \nonumber\\
&\frac{m_{i+k+1}}{3}a_{i+k+1}^2
\frac{d\Delta_{i }}{dt}\sum_{j=0}^k \Theta_{i,j}
 \hspace{1.5mm} {\rm for}\hspace{1.5mm}  k = 0,1, ... N - i - 1.\label{nomenc3}
\end{align}
Setting $i=1$ in (\ref{nomenc2}) and summing from $k=0$ to $k =N-2,$ we obtain  
\begin{align}
& \frac{d E}{dt }  - \alpha \frac{d E_{1}}{dt}= \sqrt{GM}
\frac{d\Delta_{1}}{dt}\sum_{k=0}^{N-2}\left( \frac{m_{k+2}}{3}\sqrt{a_{k+2}} \sum_{j=0}^k \Theta_{1,j}\right), \label{nomenc2a}
\end{align}
where, $E,$ is the total energy of the system and
\begin{equation}
 \alpha = \sum_{k=-1}^{N-2}\left(\frac{m_{k+2}}{m_{1}} \right) \sqrt{\frac{ a_{k+2}} {a_{1}}}\label{nomenc2aa}
\end{equation}
Similarly (\ref{nomenc3}) gives 
\begin{align}
& \frac{d J}{dt }  - \beta \frac{d J_{1}}{dt}
 = \frac{d\Delta_{1}}{dt}  \sum_{k=0}^{N-2}\left(  \frac{m_{k+2}}{3}a_{k+2}^2
\sum_{j=0}^k \Theta_{1,j}\right),\label{nomenc3a}
\end{align}
where, $J,$ is the total angular momentum  of the system and
\begin{equation}
\beta =  \sum_{k=-1}^{N-2}\left(\frac{m_{k+2}}{m_{1}} \right) \left(\frac{ a_{k+2}} {a_{1}}\right)^{2}.\label{beta}
\end{equation}
Recalling the assumption  that eccentricities are negligible,
we can combine (\ref{nomenc2a}) and (\ref{nomenc3a})
so as to eliminate $dE_1/dt$  and $dJ_1/dt$ and thus obtain
\begin{align}
&\frac{d E}{dt }  - \frac{n_1 \alpha }{\beta }\frac{d J}{dt}=\nonumber\\
&-\frac{1}{3} \frac{d\Delta_{1}}{dt}  \sum_{k=0}^{N-2}  m_{k+2} a_{k+2}^2\left(\frac{n_1 \alpha}{\beta}  - n_{k+2} \right)
\sum_{j=0}^k \Theta_{1,j}.\label{Edotsingle}
 \end{align}

For an isolated system the total angular momentum, $J,$  is conserved and we have
\begin{align}
&\frac{d\Delta_{1}}{dt} =-3\frac{d E}{dt } 
\left({ \sum_{k=0}^{N-2}  m_{k+2}a_{k+2}^2\left(\frac{n_1\alpha}{\beta}  -  n_{k+2} \right)
\sum_{j=0}^k \Theta_{1,j}}\right)^{-1}.\label{FinalLap}
 \end{align}
Thus the rate of expansion of the system, which can be shown to be positive definite, is directly proportional to the rate of energy dissipation, here associated with tidal circularization  and is thus $\propto 1/Q'$.
In a system linked together with consecutive three body Laplace resonances, the mean motion separation between any two planets is $\propto \Delta_1,$
(see ({\ref{nomenca}})). For tidal circularization the rate of energy dissipation associated with a particular planet is proportional to the square of the eccentricity,
which is $\propto 1/\Delta_1^2$
 (e.g. Papaloizou 2011).
Hence $dE/dt = (dE/dt)_0(\Delta_{1,0}/\Delta_1)^2,$ where $\Delta_{1,0}$ is the value of $\Delta_1$ at time $t=t_0$
and $(dE/dt)_0$ is the value of $dE/dt$ at that time.
 Integrating (\ref{FinalLap})  with respect to time assuming that
quantities other than the $\Delta_{i}$ may assumed to be constant,
we obtain
\begin{align}
&\Delta_{1}^3-  \Delta_{1,0}^3
 =-9(t-t_0)\Delta_{1,0}^2 \nonumber\\
& \times\left( \frac{d E}{dt } \right )_0 \left({ \sum_{k=0}^{N-2}m_{k+2}a_{k+2}^2 \left(\frac{n_1\alpha}{\beta}  -  n_{k+2} \right)
\sum_{j=0}^k \Theta_{1,j}}\right)^{-1}\label{FinalLapt}
 \end{align}
Hence for a system of arbitrary numbers of planets linked by   consecutive Laplace resonances such that the separations of neighbouring mean motions
are all proportional  to each other, 
asymptotically $\Delta_1 \propto t^{1/3}$  as has been found in other contexts where tidal dissipation operates
(e.g. Papaloizou 2011, 2015).
\subsection{Systems composed of two subsystems}\label{twosubsystems}
The TRAPPIST-1 system is such that a Laplace resonance condition  connected to the existence  of two first order resonances does not apply to planets
c , d and e. However, it does potentially apply to all other consecutive triples.
Thus the system can be regarded as a composite one in which  planets b,  c and d 
form an inner subsystem with such   
a Laplace resonance  while planets  d - h  form  an outer subsystem linked by  such Laplace resonances.

In the appendix we repeat the  analysis for a system composed of two such subsystems
with planets, $1 ,2, . . . N_1,$ forming an inner subsystem and planets, $N_1+1, N_1+2, . . .  N,$ forming an outer subsystem.
For the TRAPPIST-1 system $N_1=3.$
For such a system  it is found that   (\ref{Edotsingle})  is replaced by the equation
\begin{align}
&\frac{d E}{dt}  -\frac{n_1\alpha}{\beta}\frac{d J}{dt} =   \frac{d\Delta_{1}}{dt}\left(\sum_{k=0}^{N-2} \frac{m_{k+2}a_{k+2}^2}{3}  \left(  n_{k+2} -\frac{n_1\alpha}{\beta}\right)
\Lambda_k\right)
\label{Edotdoublet}
 \end{align}
where
\begin{align}
&\Lambda_k= \frac{ n_1-n_{k+2}}{n_1-n_2} \hspace{3mm}{\rm for} \hspace{3mm} k  <  N_1 -1,\hspace{3mm} {\rm and}
\end{align}
\begin{align}
&\Lambda_k=  \frac{n_{N_1}-n_{k+2}}{n_{N_{1}} --n_{N_1+1}}\frac{d\Delta_{N_1}}{d\Delta_1} + \frac{n_1 - n_{N_1}}{n_{1}-n_{2}}\hspace{3mm}{\rm for} \hspace{3mm} k\ge N_1-1.
\end{align}
We remark that when $\Delta_{N_1}$ and $\Delta_1$ increase monotonically with time, as found for the cases
we have  simulated,  $\Lambda_k $ is easily shown to be
positive definite and an increasing function of $k.$
This is sufficient to ensure that the quantity multiplying $d\Delta_1/dt$ in (\ref{Edotdoublet}) is negative.
It then follows that if the total angular momentum of the system is conserved, energy dissipation drives its progressive expansion.

When $\Delta_{N_1}/\Delta_1$ is constant corresponding to the situation where the ratio $(n_{i}-n_{i+1})/(n_{j}-n_{j+1})$ 
being independent of time  for $ 1 <   i  <  N-1,$   $ 1    < j   <  N-1,$
equation (\ref{Edotdoublet}) takes the same form as  (\ref{Edotsingle}) when $\Theta_{1,j}$ is expressed in terms of the mean motions.  
For more general cases, as long as both $\Delta_1,$ and $\Delta_{N_1}$ increase with time, the quantity in brackets is always positive 
so that we conclude that energy dissipation drives the expansion 
just as before and (\ref{FinalLapt}) will hold with the quantity
$\sum_{j=0}^k \Theta_{1,j}$ replaced by an appropriate mean value of $\Lambda_k.$

\section{Numerical Simulations}\label{Numsimsec}
In this section we  present  the results of numerical simulations of the TRAPPIST-1 system that incorporate tidal circularization. 
Thus the initial conditions in most
 cases  were taken to correspond to the tabulated orbital elements given by Gillon et al. (2017) and  Luger et al. (2017)
The eccentricities were assumed to be zero.  
We remark  that the actual eccentricities  cannot be strictly zero with the consequence that  we start with
a system that will not have an exact correspondence to  the actual one but will be only close to it. However, the simulations incorporate dissipative effects
which cause the system to approach a well defined evolutionary state, a feature that may mitigate the effects of not starting
with  initial conditions which exactly correspond to the  observed system.

We consider two sets of masses corresponding to models A  and B respectively.
The parameters adopted for models A  and B are listed in table \ref{table1}.
Apart from the outermost planet's mass, those for model A were taken from   Gillon et al. (2017) and Luger et al. (2017).
A value for the outermost mass is not available  so we have simply estimated it by assuming
a mean density corresponding to the earth. 
Those for  model B  were taken from  Wang et al. (2017) 
who quote a significantly smaller value for the outermost mass.
We note that the error bars are large and  we have retained  the  model  A  masses of the inner four planets
in  model  B  as differences are within them.
 We remark that the  planetary radii are the same for  models A and B.
 We  comment that tests have shown that inclusion of the outermost
planet does not affect the general form of the results for the remaining planets.
The  simulations were  by means of $N$--body calculations
 (see eg. Papaloizou \& Terquem 2001).
 From considerations  of numerical tractability  we are unable  to consider very large values of $Q'$ and very long integration times.
 Values of $Q'$ up to $\sim 12$ and integration times up to $\sim 1.5\times 10^6 y$ have been considered.
\subsection{Numerical Results}\label{Numressec}
\begin{table*}
\begin{tabular}{|c|c|c|c|c|c|c|c|c|}
Planet &  Mass&  Mean density&  Mass &  Mean density &Period & Period ratio \\
-- & $m/M_{\oplus}$&  ${\overline \rho}/{{\overline \rho}_{\oplus}}$&$m/M_{\oplus}$&${\overline \rho}/{{\overline \rho}_{\oplus}}$ &$P$&$P_i/P_{i-1}$\\
 --  &model A & model A & model B &model   B & -- &--&--  \\
1(b)&0.85&0.66 &0.85& 0.66&1.51087081& -- \\
2(c)&1.38&1.17 &1.38&1.17&2.4218233&1.602932087\\
3(d)&0.41&0.89&0.41&0.89&4.049610&1.672132727\\
4(e)&0.62&0.80&0.62&0.80&6.099615&1.506222821\\
5(f)&0.68&0.6&0.36&0.32&9.206690&1.509388707\\
6(g)&1.34&0.94 &0.57&0.40&12.35294&1.341735195\\
7(h)&0.37&1.0 &0.086&0.23&18.764&1.518990621
\end{tabular}
\caption{  The masses,  mean densities  and orbital periods in days for the planetary 
systems simulated here. The masses and mean densities for model A  were obtained  from Gillon et al. (2017)
and  Luger et al. (2017)
and those for model B   from Wang et al. (2017) (see the text for more details).
 We remark that the planetary radii adopted in both cases are the same.  
The final column  contains the period ratio with the next innermost planet where applicable. }
\label{table1}
\end{table*}
We begin by describing 
the results for  the  model A  system  with, apart from the eccentricities being set to zero,  
initial conditions from  Gillon et al. (2017) and Luger et al. (2017) and with $Q'=Q'_0=0.122127.$  For this value  of $Q',$ 
the evolution of the period ratios in the  system could be clearly exhibited
on a $10^6 y$ time scale. 
For this particular simulation circularization tides were applied only to the inner two planets.
It was found that outcomes are virtually the same when tides are applied to all of the planets consistent with
a rapid decrease of the effectiveness of tides with increasing semi-major axis.
In addition, dynamical communication between the planets is effective enough to enable
libration states to be attained for the outer planets through dissipative processes acting on the inner ones.
The value of $Q'$   adopted is unrealistically small. However, apart from slower evolution rates,  the same characteristic behaviour is found for
values of $Q'$ up to $100Q'_0.$  
The evolution  of  the  resonant angles is shown in Fig.\ref{2Qp1angles}.
These  exhibit a behaviour that is characteristic of all of the simulations starting with initial conditions close to those
given for the present system. 

The uppermost left panel of Fig. \ref{2Qp1angles} shows the behaviour of  the resonant angles
 $3\lambda_2-2\lambda_1-\varpi_2$  and  $3\lambda_2-2\lambda_1-\varpi_1$ 
that connect the innermost pair of planets. Although not in a state of libration initially these angles both rapidly attain such a state
with the libration of the former being  about $\pi$ and the latter about zero. The  amplitudes  are  quite large. A similar behaviour is found for  
the resonant angle $3\lambda_3-2\lambda_2-\varpi_2$ connecting planets $2$  and $3$ which ultimately  librates about zero and  is
illustrated in the uppermost middle panel.
On the other hand the angle $3\lambda_3-2\lambda_2-\varpi_3$ does not attain libration.
 This evolution   results in a Laplace resonance 
  being set up for planets $1, 2,$ and $3,$ but  a corresponding one 
 cannot be set up for  planets $2, 3, $ and $4$ (see the discussion at the beginning of Section \ref{SLap}).  

The resonant angles $3\lambda_4-2\lambda_3-\varpi_4$  and  $3\lambda_4-2\lambda_3-\varpi_3$ which connect planets $3$ and $4$  illustrated in
the uppermost right panel also rapidly attain a librating state analogous to those connecting $1$ and $2,$ the former librating about $\pi$ and the latter about zero, but with smaller libration amplitudes.
 A corresponding behaviour for the resonant angles $4\lambda_6-3\lambda_5-\varpi_6$  and  $4\lambda_6-3\lambda_5-\varpi_5$ 
connecting planets $5$ and $6$ is illustrated in the
 middle  panel of  the central row.
 This is also found for  the resonant angles $3\lambda_7-2\lambda_6-\varpi_7$  and  $3\lambda_7-2\lambda_6-\varpi_6$ 
illustrated in the right  panel of  the middle row.

The resonant angles $3\lambda_5-2\lambda_4-\varpi_5$  and  $3\lambda_5-2\lambda_4-\varpi_4$ that connect planets $4$ and $5$ illustrated in the left panel
of the middle row 
also librate but while the former librates about $\pi,$ the latter librates about an angle $\sim \pi/4.$
This may be connected to the fact that there are  first order resonances connecting non consecutive planets.
In particular  there  are  2:1 resonances connecting planets $4$ and $6$ and also $5$ and $7.$
The lowermost middle  panel of Fig.\ref{2Qp1angles} shows the resonant angles $2\lambda_6-\lambda_4-\varpi_6$ and  $2\lambda_6-2\lambda_4-\varpi_4$
and the lowermost right  panel shows the resonant angles $2\lambda_7-\lambda_5-\varpi_7$  and  $2\lambda_7-2\lambda_5-\varpi_5$.
In the former case the libration centres are not zero or $\pi$ while in the latter both are close to zero.
The behaviour of these angles is consistent with all triples being connected with a Laplace resonance with related first order resonances  except $2,$ $3$ and $4.$
Thus in the notation of Section \ref{SLap} where subsystems are discussed, the inner three planets can be viewed as forming
 an inner subsystem with,  $N_1=3,$
while the remaining planets form a separate outer subsystem.
 
In addition to librating in a first order 3:2  resonance, though quite distant from precise commensurability,  planets $2$ and $3$
are also close to a 5:3  resonance, though being second order this may not be expected to play a significant role on account of the small eccentricities.
 However, the resonant angle  $5\lambda_3-2\lambda_2-\varpi_3$ illustrated 
in lowermost left  panel of Fig. \ref{2Qp1angles} indicates periods of large amplitude
 libration indicating that this resonance may be involved in the dynamics.

The time evolution of the   period ratios of consecutive planets and orbital  eccentricities
is shown in Fig. \ref{2Qp1periodratios}.
The period ratios  $P_2/P_1,$
$P_3/P_2,$ $P_4/P_3,$ $P_5/P_4,$ $P_6/P_5$ and 
  $P_7/P_6$  all tend to increase with time in the same way.
This is consistent with the expansion of the system discussed in Section \ref{SLap}
where it is argued that the evolution time scales as $Q'.$
For the simulation discussed here with $Q'=Q'_0$, period ratios depart significantly from initial values after $\sim 2\times 10^6 y,$ 
indicating values of  $Q'$ in the range   $10^{2-3}$  are required if the system is to remain near the present state over $Gy$ time scales.
The evolution of the semi-major axes of  all of the  planets is illustrated in the lowermost left panel.  We plot
$\log (a_i/a_{i0}) -0.001(i-1) $ for $i=1-7,$ where $a_{i0}$ is the initial value of the semi-major axis of planet $i.$
The shift  $-0.001(i-1)$ is applied simply to allow visibility of all curves. 
The innermost  four planets  migrate inwards and the  outermost three outwards as consistent with a general  expansion 
of the whole  system  while conserving
total angular momentum as described in Section \ref{SLap}.
In addition planet $3$ migrates inwards such that both $\Delta_{N_1}$ as well as $\Delta_1$ increase with time as assumed in Section \ref{SLap}.
The eccentricities of all the planets are illustrated in the lowermost middle and right panels.   While 
remaining small and $< 0.01,$
they decrease secularly with time. This is a consequence of the increasing period ratios which move planet pairs further away from resonance. 

\subsection{Dependence on $Q'$}\label{Qdep}
In order to  investigate the dependence on, $Q',$
we present the results of a simulation with the same parameters as the preceding one except that
$Q'=100Q'_0$   but with tides now  allowed to act on all of the planets.
 The evolution   of the  resonant angles is shown in Fig.  \ref{2Qp100angles}. This plots the same quantities that are shown
in Fig. \ref{2Qp1angles} for the case with $Q'=Q'_0,$ so that the results may be directly compared.
We see that the resonant angles connecting planets $1$ and $2,$ those connecting  planets $2$ and $3,$ those connecting $3$ and $4,$
those connecting $5$ and $6$ and those connecting $6$ and $7,$ apart from taking longer to establish clear libration, show very similar behaviour.
The increase in the  time required to establish libration can be up to two orders of magnitude but is not uniform and the libration amplitudes are somewhat larger.
The resonant angles $3\lambda_5-2\lambda_4-\varpi_5$  and  $3\lambda_5-2\lambda_4-\varpi_4$ that connect planets $4$ and $5$ illustrated in the left panel
of the middle row  librate as before  but with the latter now librating  about an angle $\sim  7\pi/4.$ 
The resonant angles associated with the $2:1$ resonances
connecting non consecutive planets shown in the lowermost panels also show shifts of libration centre as compared to the  previous simulation.   

The behaviour of the  period ratios  and eccentricities  is illustrated 
in Fig. \ref{2Qp100periodratios} and it can be directly compared to that for the previous simulation
illustrated in Fig. \ref{2Qp1periodratios}. It will be seen that with the much larger $Q',$ the period ratios do not show a noticeable secular increase
and the system as a whole does not show a measurable expansion.
This is because the eccentricity damping time scale is increased by a factor of a hundred in this case resulting in a
corresponding increase in the evolution time scale for the period ratios which accordingly do not show significant
evolution over the time shown here.
However, it is possible to detect period ratio evolution
for values of $Q'$ up to $10Q'_0$ over this time period. Simulations with  $Q'=Q'_0$ and $Q'= 10Q'_0$ were used to verify
that the rate of period ratio evolution was indeed  $\propto 1/Q'$ as expected from the discussion of Section \ref{SLap}.

\subsection{Effect of changing the masses}\label{masseffect}
 To investigate the effect of changing the masses we present the results of two simulations with the same parameters as the preceding two, except that 
masses corresponding to model B  were adopted and tides were applied to all planets. 
The evolution of the  resonant angles  for the case   $Q' =Q'_0$ is illustrated in   Fig. \ref{2Qp1newpassesangles} 
and  for the case $Q' = 100Q'_0$  is illustrated in Fig. \ref{2Qp100newpassesangles}.
The period ratios and eccentricities for these simulations are shown in 
Figs. \ref{2Qp1newpassesperiodratios} and \ref{2Qp100newpassesperiodratios} respectively. 
A comparison with the preceding simulations for which  masses appropriate to model A  were used
indicates similar behaviour for the resonant angles. However,  libration amplitudes are larger and  it can take longer to attain  librating states.
This is particularly the case for angles associated with the outermost planet  which seems in general  to show less stable behaviour
on account of its low mass.
The behaviour of the period ratios and eccentricities is similar to the case for which the masses were appropriate to model  A.
 The evolution of the semi-major axes in the case with $Q' = Q'_0$ is such that the inner three planets migrate inwards,
for the fourth planet it remains on average nearly constant, while the outer three move outwards. On account of its small mass,
the outermost planet moves outwards significantly more rapidly than the simulation with case  A masses and $Q' = Q'_0.$

\subsection{Interaction of the inner and outer subsystems}\label{Intsub}
Apart from planets 2 and 3,
consecutive pairs of planets in the TRAPPIST-1 system are linked by first order resonances  for which both associated angles librate and are configured such that
Laplace resonance conditions connecting  consecutive triples may hold in a time average sense for every case except planets $2,$ $3$ and $4.$
Furthermore the deviations from exact commensurability are much greater for the innermost three. 
This indicates that the system  may be viewed  as being comprised of two subsystems of the type considered in Section \ref{SLap}
that are completely linked by such  Laplace resonances. The innermost one consisting of the first three planets and the outermost one of the remainder.

In this section we provide a preliminary exploration of  the possibility that
a  subsystem consisting of either the inner two or three planets formed separately from the rest 
  and subsequently interacted with them.
In particular we suppose that the inner subsystem expanded to interact with the outer subsystem that was  essentially stationary, though  we remark that there are variations of this
such as where the outer subsystem migrates inwards while the inner one remains stationary that are likely to lead to similar conclusions.  
To explore this scenario, we present the results of  simulations for 
which either the system was set up with the inner three planets 
in a Laplace resonances but with smaller initial semi-major axes, or
for which only the inner two planets were similarly set up in a closer 3:2 resonance. 
The separate subsystems then expand with much weakened coupling to the corresponding outer system.
Parameters are set such that the systems begin to interact more strongly when conditions
corresponding to the current observed system are approached
and we investigate whether the interaction leads to the system eventually
evolving in a similar way to the one initiated with initial conditions corresponding to the observed system.

We do not consider the history of the outer system, noting that it could be in a different configuration to that corresponding to
the same planets in the  observed system. For example if it migrated inwards
until reaching a protoplanetary disc edge that was progressively moving outwards, 
 allowing the inner system that formed earlier to have moved closer to the star,
 the resonances might have been closer to commensurability.


In the first simulation presented in this Section,   tidal circularization is applied 
 only to the innermost two planets,  with $Q'=Q'_0,$   and the masses  corresponded to model A. 
The system was set up as for the observed system with the difference that 
 the initial  periods of planets $1$, $2$ and $3$ were set to   $ 1.52647745 $ days,      
$2.422519714 $  days and $3.97777737 $  days respectively. As in all other cases the initial eccentricities were set to zero.
 These correspond to the period ratios $P_2/P_1 = 1.587,$ and $P_3/P_2 = 1.642.$
These are smaller than for the  observed system as they correspond to  a situation where the inner subsystem of planets $1$, $2$ and $3$ is in a Laplace resonance but
is less expanded. Note that although planet $3$ is at a reduced period as compared to the observed system, planet $1$ has a longer period.
The evolution of the resonant angles for this simulation is illustrated  in Fig. \ref{2QP2angles}.
 and the evolution of the period ratios  is illustrated in Fig. \ref{2QP2periodratios}.
 
 The evolution can be split into two phases. Before the subsystems  start to interact significantly at time $\sim 4.5\times 10^5 y,$ the inner system
 evolves like a separate system of $N = 3$ planets linked by a Laplace resonance as described  in Section \ref{SLap} 
 and equations (\ref{FinalLap}) and  (\ref{FinalLapt}) apply (see also Papaloizou 2015).
 From Fig. \ref{2QP2periodratios}, the estimated time for this subsystem to separate to attain current period ratios starting  from 3:2
 commensurabilites is $\sim 10^6 y,$ again indicating that values of $Q' \sim 10^{2-3}$ are needed in order to obtain evolution
 on $Gy$ time scales.
   
 During the first  phase the outer planetary system is only weakly coupled to the inner one. 
 Thus the associated resonant angles show more erratic behaviour than the observed system
 though many show libration while the outer period ratios do not show a secular increase. After an initial  interaction phase is complete at $t \sim  6\times 10^5 y$
 the entire system enters an evolutionary phase resembling that obtained when one starts with parameters corresponding to the observed system
 with  resonant angles showing stable librations and the period ratios secularly increasing.
In order to investigate the behaviour of orbits with  initial conditions in the neighbourhood  
 of those implemented above we considered additional simulations 
for which the orbital phases of all planets were randomised. Out of fourteen  such  realisations,
 four  evolved in the same way as described above, the Laplace resonance being  broken
in the other cases with  subsequent
 evolution causing the period ratio, $P_3/P_2$ to decrease and thus move away from the value appropriate to
the observed system.
This indicates that in spite of some fragility,
the form of evolution that approaches that found for initial parameters corresponding to the observed system is not that of an isolated case.

 In the second simulation presented in this Section, the set up was as for the observed  system
except that the initial periods of the inner two planets  were  
taken to be  $1.60239588$ days  and $2.414794275$ days. Other parameters were the same as  those for the
the first simulation described in this Section. This corresponds to an initial period ratio of $1.5069$
for the innermost pair while planets $2$ and $3$  are slightly more separated than in the observed system with a period ratio
of  $1.677.$ Thus the inner pair starts in a close 3:2 resonance with the second planet just wide of
a  5:3  resonance with the third. In this case the inner pair, assumed to have formed separately,   a possibility  indicated in  Section \ref{PSEC} below,  separates 
as a result of tidal interaction  and thus interacts  with the outer planets.

The evolution of the resonant angles for this simulation is illustrated in Fig. \ref{2Qp2anglesmig}
and the evolution of the period ratios 
is illustrated in Fig. \ref{2QP2periodratiosmig}.
As the inner two planets separate, the period ratio between the second and third planets
decreases and a 5:3 resonance is formed. This couples them to the outer planets
transferring angular momentum to them and increasing their eccentricities.
The latter phenomenon results in decreasing period ratios, 
just as decreasing eccentricities leads to them increasing.  
After about one megayear the period ratio of the innermost
pair has increased enough to enable the  inner three planets to settle into a Laplace resonance, when the required  condition on the mean motions
is satisfied. Subsequently
the evolution approaches that of the observed system for which the  eccentricities decrease 
and the period ratios increase secularly with time. 

To establish that orbits close in phase space to the one illustrated above undergo
the same type of evolution, we interrupted it at a late stage  but well before attaining the final state corresponding to the observed system 
and randomised the orbital phases of all planets for ten different realisations. This caused the 5:3 resonance between planets 2 and 3 to be broken. However after a brief
relaxation phase it was reestablished and their evolution approach the same form as in the uninterrupted case.


\begin{figure*}
\begin{center}
\vspace{1cm}
\hspace{-0cm}\includegraphics[width=0.45\textwidth]{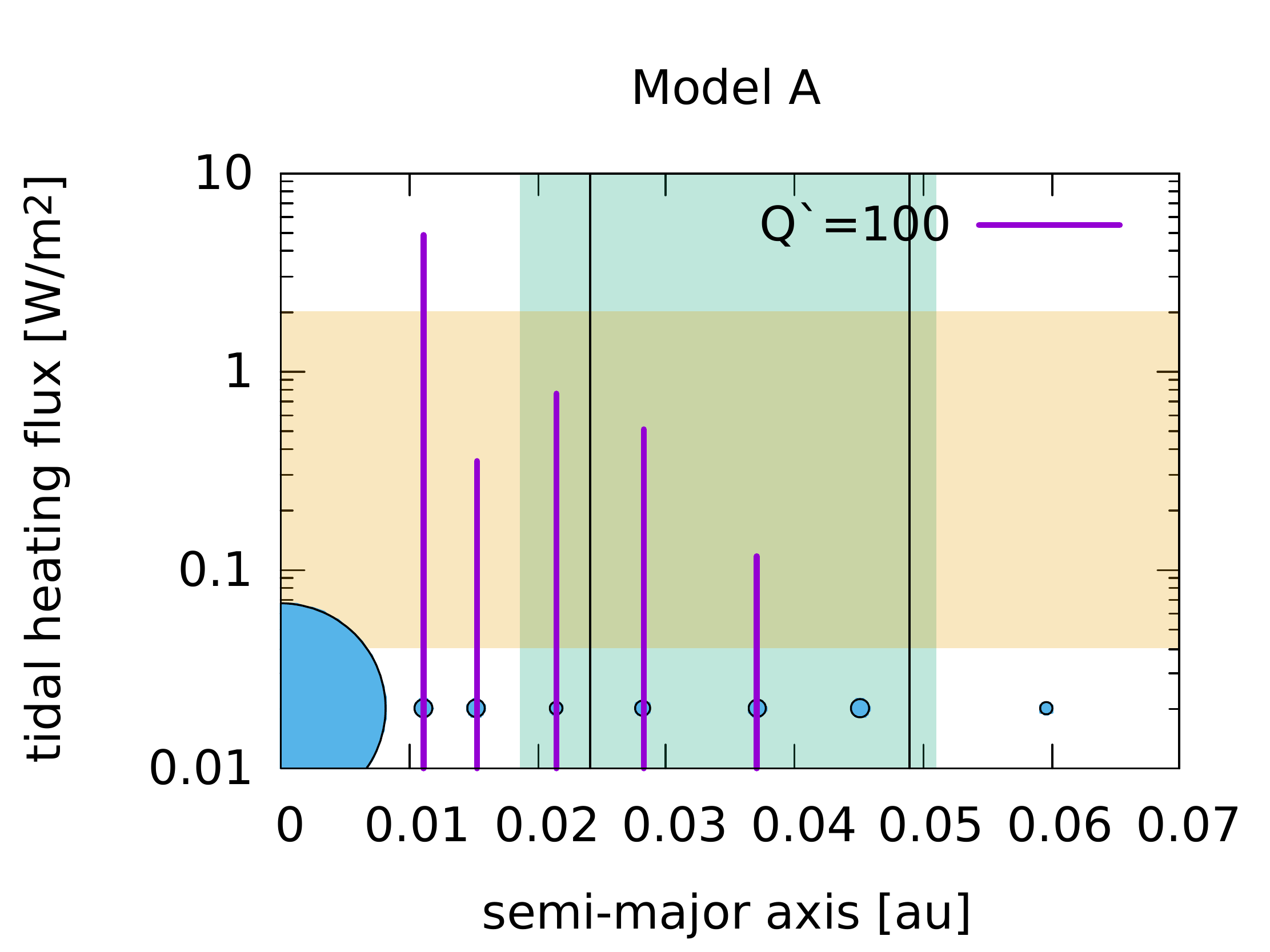} 
\hspace{1.cm}\includegraphics[width=0.45\textwidth]{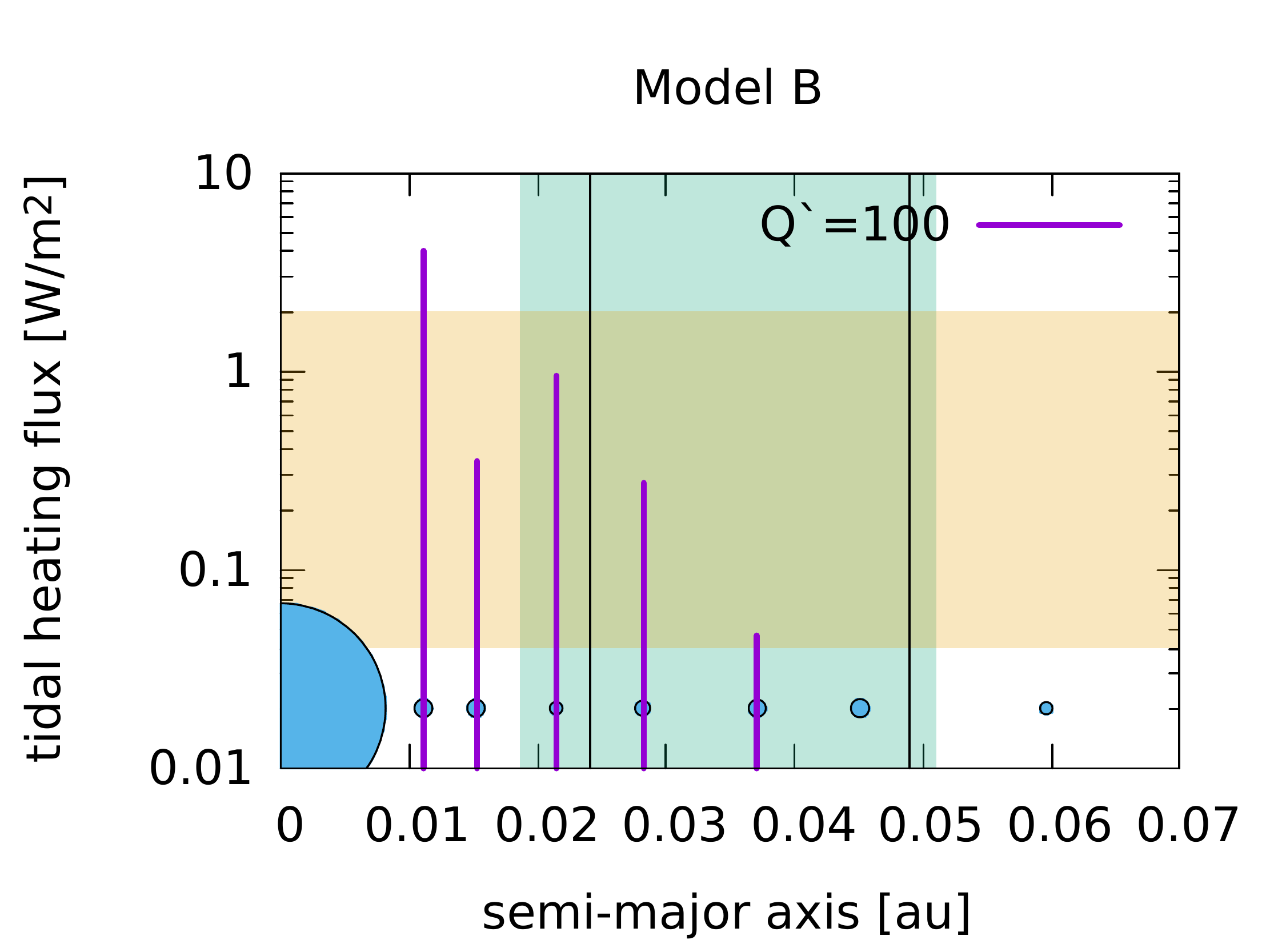} 
\vspace{1cm}
\hspace{-0cm}\includegraphics[width=0.45\textwidth]{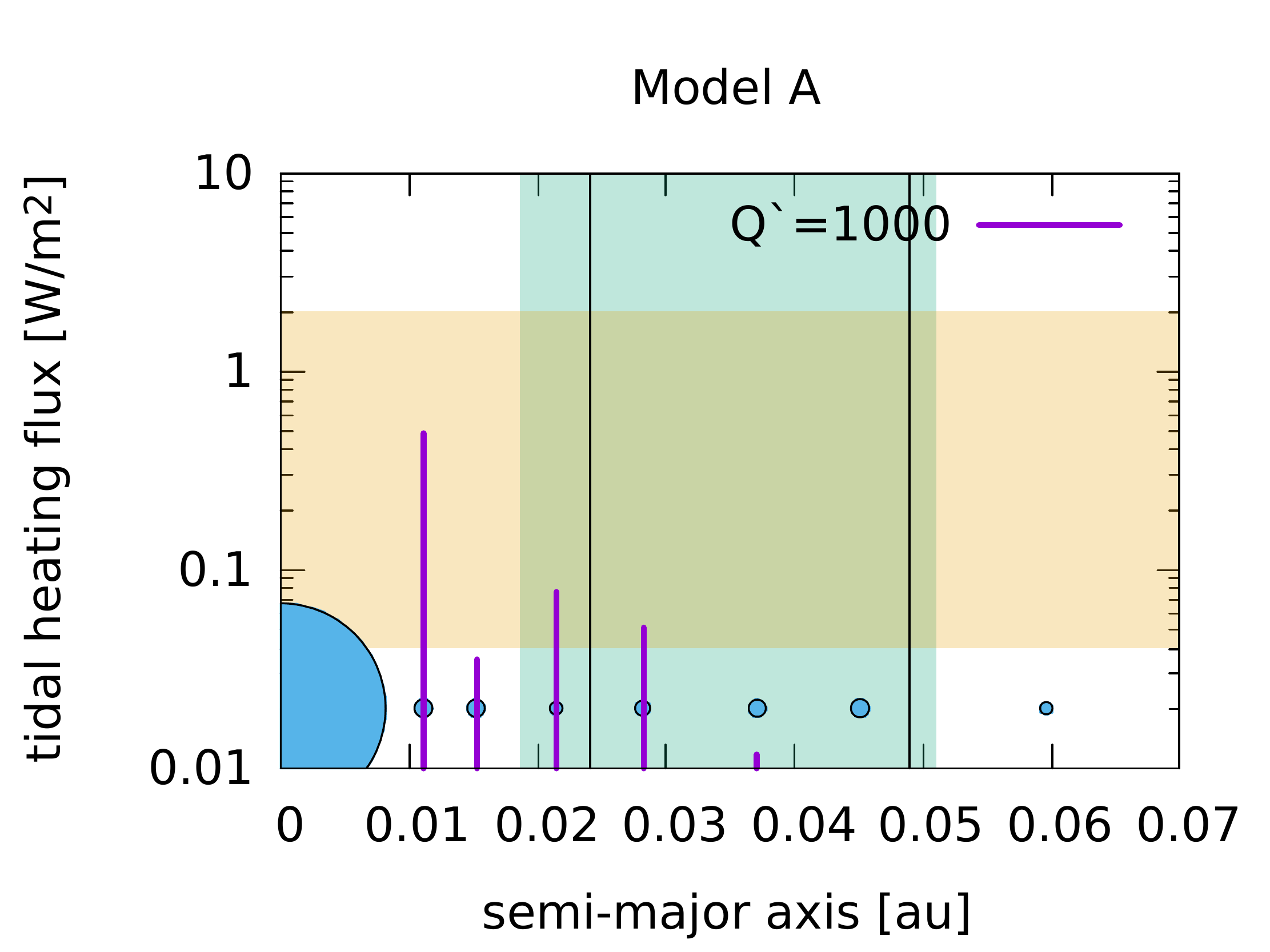} 
\hspace{1.cm}\includegraphics[width=0.45\textwidth]{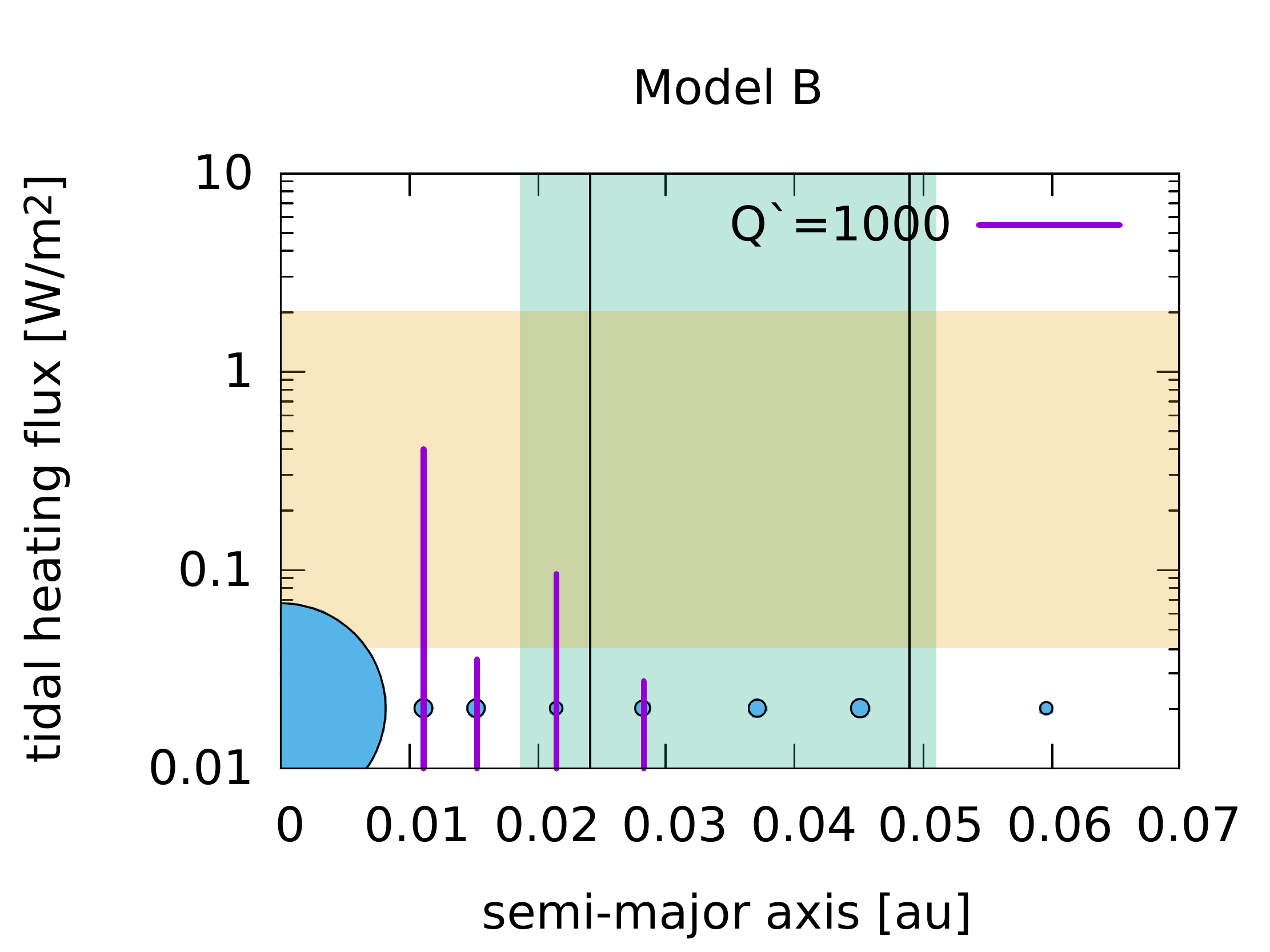} 
\end{center}
 \caption{
The TRAPPIST-1 system together with its habitable zones (insolation  
zone in green  and tidal zone in yellow). The violet impulses show
the amount of the tidal heating flux produced by each planet. The tidal 
heating flux in the range  from 0.04 till 2 $W/m^2$ favours plate
tectonics. The results are shown for two different sets of planets'
masses: model A (left) and model B (right) and for two different
values of $Q'$: 100 (upper panels) and 1000 (lower panel).
The borders of the conservative
habitable zone are marked by two vertical black lines.
\label{habi}}
\end{figure*}

 \section{ Maintenance of commensurabilities  when a  protoplanetary disc is present and the effect of disc dispersal
  on  resonant chains}\label{disceffect}

Tamayo et al. (2017)
have suggested that the chain of resonances in the TRAPPIST-1 system
may have formed while the planets were migrating through the disc
towards the central star.  In their simulations, all the planets are
subject to eccentricity damping due to interaction with the disc, but
only the outermost planet~7 feels a direct migration force at the
beginning of the calculation.  After it captures planet~6 into a
resonance, the pair migrates in together, which leads to capture of
planet~5 and so on, until all the planets are in resonances.  It is
found that by adjusting the ratio of the migration and eccentricity
damping timescales the observed chain of commensurabilities can be
reproduced.  However, an issue with this is  that all the planets
should be subject to direct migration forces from the beginning of the
calculation.  The  interaction with the disc that leads to
eccentricity damping also in general  produces   migration  torques unless
 there is a  balance between Lindblad and corotation torques for each planet
apart from the outermost one. It would seem unlikely that this could be maintained
at  all  disc locations that they move through.
The migration prescription is not an arbitrary function but should be
related to a disc and have an appropriate dependence on the planet mass
(see eg. Papaloizou 2016).

As shown below, planet~2 is expected to migrate faster than planet~3
because, being significantly more massive, it interacts more strongly
with the disc.  As a consequence, it cannot maintain a resonance with
planet~3 while the system migrates through the disc.  Therefore
planets~1 and~2 reach the disc inner parts before the other planets,
and the chain of resonances can be established only after all the
planets have arrived in the central parts and terminate migration
there.  Termination may be due either to the planets entering an inner
cavity, or to the innermost planet stalling just beyond the disc's
inner edge.

 Masset et al. (2006)
have indeed suggested that cores could be
trapped at the edge of the disc, rather than penetrating inside the
cavity, through the operation  of  corotation torques.
Whether that
happens or not is not clear, as it may depend on the level of MHD
turbulence at the inner edge.  Also, if the disc's inner edge moves
out because of X--ray photo-evaporation, the surface density of gas in
the vicinity of the innermost planet would decrease to zero.  In that
case, the disc cannot transfer enough angular momentum to the planet
for it to stay coupled and move outward with the retreating inner edge.  Given
the uncertainties, here we consider both  possibilities, with either the
planets entering the cavity, or stalling just  beyond the disc's inner
edge.

Ormel, Liu and Schoonenberg (2017)
have considered a model in which the planets in the TRAPPIST-1 system
move inward and stall at the disc's inner edge.  They assume that
consecutive pairs among planets~1 to~5 are initially in 3:2 MMRs, and
they investigate the conditions under which these can be broken for
the two innermost pairs, which are observed to be close to 8:5 and 5:3
commensurabilities.  Ormel et al. (2017) assume that the disc's inner
edge moves out due to the expansion of the magnetospheric cavity, and
that the planets decouple from the disc one after the other, starting
with the innermost one.  As decoupling happens on a timescale which
depends on the planet's mass, a period of divergent migration may take
place when a planet has decoupled and the next outermost one is
temporarily moving outward with the disc.  By adjusting the expansion
rate of the disc's inner edge, Ormel et al. (2017) have found that the
MMRs for the two innermost pairs could be broken.  It is not clear
however how these planets would then attain the observed period ratios
that are consistent with there being a three body Laplace resonance
between the inner three planets as indicated in our discussion in
Sections \ref{SLap}, \ref{Numressec} and \ref{Intsub}.

Note that in Section \ref{SLap} we have presented an alternative
scenario where the large separations from strict 3:2
commensurabilities for the two innermost pairs can in fact be produced
through tidal interaction with the central star.

In this section, we consider a system of 6 planets which start close
to the disc's inner edge and which either subsequently enter an inner
cavity, or stall just beyond the disc's inner edge in a resonant
chain.  We ignore the outermost planet as its mass is not well
constrained, but this does not affect the conclusions of our study.
The goal of these simulations is to study whether commensurabilities
can be maintained during the evolution of the system and the disc.

\subsection{Planetary systems that enter  the cavity}\label{PSEC}

We set up the 6 inner planets of the TRAPPIST-1 system on circular
orbits in a disc with inner edge at $r \equiv R_{\rm in} = 0.05$~AU.
The value of $R_{\rm in}$ is chosen so that the innermost planet is
close to its current location after all the planets have penetrated
inside the cavity.  The orbit of the innermost planet is set with a
semi-major axis $a_1=0.055$~AU.  We consider two different sets of
initial conditions, where planets~2 to~6 are started at distances such
that $P_2/P_1=8/5$, $P_3/P_2=5/3$ (case A) or $P_2/P_1=3/2$,
$P_3/P_2=3/2$ (case B) while $P_4/P_3=P_5/P_4=3/2$ and $P_6/P_5=4/3$
in both cases.  Case~A is close to the observed system. The motivation
for considering case~B is that this configuration would be easier to
obtain through migration as it involves only first order resonance,
and the configuration of the final chain of MMRs may have resulted
from later tidal interaction with the star (see, e.g., Section
\ref{Intsub}).  We then solve the equations of motion~(\ref{emot}) for
each planet where accelerations due to tidal interaction with the disc
are given by equations~(\ref{Mui}) and~(\ref{Gammai}).  In the regime
of inward type--I migration, Papaloizou \& Larwood (2000) have shown
that the migration and eccentricity damping timescales can be written
as:

\begin{equation}
\begin{split}
  t_{mig,i}  = 146.0  \; \left[ 1+ \left( \frac{ e_i }{1.3 H/r}\right)^5 
\right]
  \left[ 1- \left( \frac{ e_i }{1.1 H/r}\right)^4 \right]^{-1} \\
  \left( \frac{H/r}{0.05}
  \right)^2 \;
  \frac{{\rm M}_\odot}{M_d} \;
  \frac{{\rm M}_\oplus}{m_i} \; \frac{a_i}{{\rm 1~au}}
  \left(  \frac{{\rm M}_\odot} {M} \right)^{1/2}   \; \; \;  {\rm years} ,
\label{tm}
\end{split}
\end{equation}

and

\begin{equation}
\begin{split}  
t_{e,i}^d = 0.362 \; \left[ 1+ 0.25 \left( \frac{ e_i }{H/r}\right)^3
  \right] \;  \left( \frac{H/r}{0.05}
  \right)^4  \\
  \frac{{\rm M}_\odot}{M_d} \; \frac{{\rm M}_\oplus}{m_i} \;
  \frac{a_i}{{\rm 1~au}}  
\left( \frac{{\rm M}_\odot}{M} \right)^{1/2} \; \; \; {\rm years} ,
\label{te}
\end{split}
\end{equation}

\noindent Here $H/r$ is the disk aspect ratio, which is taken to be
0.05, and $M_d$ is the disk mass contained within 5~au.  The equations
 assume that the disc surface mass density
$\propto r^{-3/2}$.  The eccentricity damping timescale taking into account
both tidal interaction with the  star and disc interactions can be calculated through
$1/t_{e,i}=1/t_{e,i}^d +1/t_{e,i}^s$, where $t_{e,i}^s$ is given by
equation~(\ref{teccs}).

Masses derived for the $\sim$ 1~Myr old discs in the Ophiuchus
star-forming region are on the order of 0.01~${\rm M}_\odot$ within
50~AU (Andrews et al. 2010).
Most of the stars in this study have a
mass between 0.6 and 2~${\rm M}_\odot$. As these discs are best fitted
with a surface mass density $\propto r^{-1}$, this implies that the
mass within 5~AU is on the order of $10^{-3}$~${\rm M}_\odot$ after
1~Myr.  In addition, observations and modelling of discs around
T~Tauri stars in the Taurus and Chamaelon~I molecular cloud complex
suggest that the disc's mass decreases by a factor~10 during the first
Myr of evolution (Hartmann et al. 1998).
If we assume that the disc
mass scales with the central mass,
this suggests that the mass within 5~AU
in the disc surrounding TRAPPIST-1 would have been roughly between
$10^{-4}$ and $10^{-3}$~${\rm M}_\odot$ during the first Myr of
evolution.

Figure~\ref{fig:mig} shows the evolution of the system for case~A and
$M_d=3.5 \times 10^{-4}$~${\rm M}_\odot$.  Both $Q'=100$ and $Q'=1000$
are considered.  The evolution of the semi-major axes and period
ratios are only shown for $Q'=100$, but the plots for $Q'=1000$ are
very similar.  As the planets migrate inward, they get closer to each
other so that the eccentricities increase.  They subsequently decrease
in the case $Q'=100$ when the innermost planet reaches radii small
enough that tidal interaction with the star becomes important.  Note
that similar eccentricities are obtained when the mass of the disc is
halved.  The 3:2 and 4:3 MMRs between planets~4 and~5 and between
planets~5 and~6, respectively, that were present initially, are
maintained during the whole duration of the simulation.  Planets~2
and~3 were initially in a 5:3 MMR, but as planet~2 is significantly
more massive it migrates faster than planet~3 and the MMR is
destroyed.  Only after planet~2 penetrates into the cavity can
planet~3 catch up and at that point a 2:1 MMR is established.  In
addition planets~1 and~2 establish a 3:2 resonance that is maintained
for the duration of the simulation.  Planets~3 and~4 were initially in
a 3:2 MMR, but this gets temporarily destroyed as planet~3 is dragged
behind planet~2 at the beginning of the simulation.  After some time
though, the 3:2 MMR is re--established.  After the phase of
readjustment of the MMRs, all the resonant angles and/or angles
between the apsidal lines for pairs of consecutive planets librate
around some fixed value, and departure from exact commensurability is
at most of a few tenths of a percent.

Similar results are obtained in case~B, except that planets~1 and~2,
initially set  up in a 3:2 MMR, remain in this resonance for the whole
duration of the simulation.  In this case again, planets~2 and~3
end up in a 2:1 MMR,  even though they started in a 3:2 MMR.

The calculations described here show that the observed chain of
MMRs cannot be established as planets migrate through the disc because
planet~2 migrates too fast for a resonance with planet~3 to be
maintained.  This indicates that planets~1 and~2 must have reached the
disc's inner parts before the rest of the system, and MMRs were later
established when the outer planets joined them.  If planets have
indeed been able to penetrate inside a cavity, rather than stall
beyond the disc's inner edge, large eccentricities must have been
excited during the migration process.  As the upper limit on the
observed eccentricities is of a few percent, stellar tides must have
been efficient enough to damp the eccentricities produced by the
migration process.


\subsection{Planets stalled beyond the disc's inner edge}

We now study the evolution of the system when the planets stall beyond
the disc's inner edge.  We set up the 6~planets on circular orbits at
the distance from the star where they are currently observed, and with
consecutive planets in MMRs.  Here again we consider both cases~A
and~B.  We assume that the disc's inner edge lies just interior to the
location of the innermost orbit.  The tidal torque from this edge
prevents the innermost planet from migrating inside the cavity.  This
torque is passed on from one planet to the next, and as a result all
the planets are prevented from migrating inwards.  However, as they
are still embedded in the disc they are still subject to eccentricity
damping.  We therefore adopt the same governing equations as in the
previous Section with the same eccentricity damping timescale,
$t_{e, i}^d,$ but ignoring $t_{mig,i}$.  We also ignore $t_{e,i}^s$
here as it is much longer than $t_{e, i}^d$ so that
$t_{e,i}=t_{e, i}^d$ as long as the disc is present.  In the initial
set up, consecutive planets are in MMRs, but resonant angles are not
librating.  Like in the case where stellar tides are acting and
described in Section~\ref{SLap}, libration
of the resonant angles is obtained after some transition period.  Because
energy is dissipated through damping of the eccentricities at constant
angular momentum, the system expands and there is an increasing
movement away from exact commensurabilities, in the same way as when
eccentricities are damped by tidal interaction with the star.
However, as here eccentricity damping happens on a shorter timescale
and acts on all the planets simultaneously, the departure rate from
exact commensurabilities is larger.  This departure rate increases
sharply at the beginning of the simulations, which may be due to the
initial set up, in which the resonant angles are not librating, and
more slowly later on, after the planets are dynamically coupled.
After the departure rate starts increasing more slowly, we disperse
the disc on a prescribed timescale.  It has been shown in previous
studies (Cossou et al. 2014, Coleman \& Nelson 2016, Terquem 2017)
that MMRs can be destroyed when the disc disperses, and therefore it
is important to check under which conditions can the MMRs in the
TRAPPIST-1 system  be maintained.  In the simulations described
below, dissipation of the disc is modelled by decreasing its mass
linearly down to zero starting at time $t=t_{\rm dis}$ and over a
timescale $t_{\rm evol}$.

In figure~(\ref{fig:stal}), we show the evolution of the departure
from exact commensurability for all the pairs of consecutive planets.
We define $\delta_{i,i+1} \equiv 100(P_{i+1}/P_{i}-r_{i,i+1})$, where
$r_{i,i+1}$ is the initial value of $P_{i+1}/P_{i} $.  So
$\delta_{i,i+1}=1$ indicates that the departure from exact
commensurability for planets~$i$ and~$i+1$ is 1\%.  For
$M_d=3.5 \times 10^{-4}$~${\rm M}_\odot$, in both cases~A and~B, the
departure becomes larger than 1\% for several pairs of planets, even
if we discard the initial sharp rise of $\delta_{i,i+1} $ which may be
due to the initial set up.  Note however that departure from exact
commensurability is larger in case~B, i.e. when all the planets start
closer to first order resonances.

As the damping timescale is shorter when the disc mass is larger, at a
given time departures from exact commensurability are larger for more
massive discs.  For the departure to be below 1\% for all pairs of
planets the mass of the disc has to  be smaller than
$3.5 \times 10^{-5}$~${\rm M}_\odot$ in case~A.  We have found that in
case~B, even for $M_d$ as small as $10^{-5}$~${\rm M}_\odot$, the
departure from commensurability is still 1\% for several pairs of
planets.  Dispersal of the disc decreases this, but the effect is
significant only for larger departures from commensurability, and in
that case it is not enough to bring departures from commensurability
below 1\%.  The timescale $t_{\rm evol}$ on which the disc disperses
does not affect the results significantly.  The longer the system
stays in the disc before it disperses, the larger the departures from
commensurability become.  Recall that action of stellar tides, which
we have discussed but not included here, produces the same type of
evolution.

We have found that libration of the resonant angles and of the angles
between apsidal lines associated with consecutive pairs of planets,
when it takes place, is maintained as long as the disc is present,
even when deviations from commensurabilities increase.  However, it is
destroyed when the disc is removed on a timescale shorter than
$10^6$~years.  For the resonant angles this happens rather suddenly
when there is hardly any disc left, whereas for the angles between
apsidal lines it happens a bit more progressively.  When
$t_{\rm evol}=10^6$~years, libration can be maintained.  However, we
recall that if destroyed by disc dispersal, as we have shown above,
libration can be restored later as a result of tidal interaction with
the star.

\section{Consequences for habitability}\label{Habitability}

The TRAPPIST-1 system is a promising target for an attempt to
characterise  potentially habitable planets. 
This is because its star is an ultra-cool dwarf of the spectral 
type M$8V$, located only 12 pc 
away from the Sun. Its seven confirmed planets have masses in the 
terrestrial mass range.  Four of them receive the appropriate amount
of stellar flux to enable liquid water to exist  on their surfaces 
on optimistic estimates, depending on the atmospheric characteristics
and precise orbital configurations of the planets (Gillon et al. 2017, O'Malley-James
\& Kaltenegger 2017).  

It is generally expected that a habitable planet (for life as we know it) 
should fulfill  a number of requirements.  Such  a planet should orbit 
a long-lived star, which shines steadily for millions of years. The stellar 
flux needs to be sufficient to enable the existence of liquid water on the planet's 
surface. The mass of the planet also matters for holding a  substantial
life-supporting atmosphere. Additional 
important demands for habitability are connected with  plate tectonic
activity and a protective magnetic field.
All of these properties have been already discussed 
in the context of TRAPPIST-1 system. The  requirement of
orbiting a long-lived star is easily satisfied as a low-mass M-dwarf
is an extremely long-lived and therefore at least in principle there is a plenty
of time for the biological evolution of its planets. However, in order to
determine a potential habitability it is necessary
to know  the present age of  TRAPPIST-1. Unfortunately,
this is  still poorly constrained.

As reported in Filippazzo et al. (2015) TRAPPIST-1  should be older than 0.5 $Gy$.
However, Luger et al. (2017) argue that  its age is in the
broad range of  $3 - 8 Gy$  while  Burgasser \& Mamajek (2017) claim that the age
of the star is  $7.6 \pm 2.2$ $Gy,$ being close to the upper limit of this  range.
Other properties of the star, for example their  UV/EUV/X-ray fluxes, seem
to indicate the younger age (Bourrier et al. 2017, O'Malley-James \& Kaltenegger 2017).
More studies are needed  to resolve this  issue definitively.
 Accordingly  for the purpose of our
present investigation we shall only  assume  that  the  age of TRAPPIST-1  exceeds  $0.5 Gy.$

The requirement for habitability  associated  with the stellar flux and the existence of  liquid water
has been discussed by many authors  
(eg.  Gillon et al. 2016,  Luger et al. 2017, Wolf 2017 and Turbet et al. 2017).
Tidal heating which can be important for plate tectonics activity has been considered
by  Gillon et al. (2017) and  Luger et al. (2017) and has been discussed in detail in Turbet et al. (2017).
 A  protective  magnetic field is a requirement for habitability. 
M-dwarfs have not only extremely long lifetimes, but also very high level of
magnetic activity, particularly when they are young. The X-ray/EUV emission of  TRAPPIST-1
has been studied by Wheatley et al. (2017) and Bourrier et al. (2017). Garraffo et al. (2017)
model the space weather conditions of the planets around TRAPPIST-1.


In this section  we  evaluate  the   rates  of
internal heat generation by  the tidal forces that we found to be  consistent  with 
for the  current orbital configuration  following the modelling carried out
in Sections \ref{Numressec} and \ref{Intsub}.  This  is  potentially relevant to
climate-regulating plate tectonics.

The rate of energy dissipation produced in  a planet
as a result of tidal heating
is  given by  
equation~(\ref{Edissip}) with the circularization time scale taken from equation~(\ref{teccs}). 
For the  preliminary  estimates  made here, we note that
for the most part
 we are making  the assumption that $Q'$
is the same for all planets in the system, 
while remarking that its adoption is uncertain and  that it 
should be relaxed in a more complete study. 
Furthermore  we  use simple scalings to extrapolate results to larger values of $Q'$
than those adopted in the simulations.

The surface effects of tidal heating on the potential biosphere
of  planet $i$ can be expressed in terms of the tidal heating flux
through
\begin{equation}
h_i = \frac{(dE_i/dt)}{4\pi R_i^2}
\end{equation}
where $R_i$ is the radius of planet $i$.
We assume after Barnes et al. (2009b) that a tidal heating flux in 
the range of 0.04-2 $W/m^2$ favours  plate tectonic activity at  a
level  conducive to life  and accordingly, this  
can be  described  as defining a "tidal habitable zone".

 We consider  both
sets of planetary masses: models A and B. 
From the simulations of the evolution of the systems 
we obtain time averaged values of the eccentricities
of each planet  and then use equation~(\ref{Edissip})
to determine that rate of energy dissipation.
Once the system of all seven planets has settled down
with orbital elements close to their observed values, the time averaged eccentricities
are found not to depend on $Q'$ for  $Q' > 100Q'_0.$
This has the consequence that in this regime,  the dissipation rate is 
expected to be  $\propto 1/Q'.$

In Fig. \ref{habi} we illustrate the tidal heating flux for the seven
planets of TRAPPIST-1 for $Q' = 100,$ and $Q' =1000.$
The yellow region denotes the tidal habitable
zone, where the plate tectonics may be active. The 
insolation optimistic habitable zone is the
green region and the conservative zone extends between the two
black vertical lines (Kopparapu et al., 2013; Kopparapu et al. 2014).


 It can be clearly seen that for $Q'=100$  there are two planets, namely
TRAPPIST-1 e and f, that are located not only in the conservative habitable
zone but also in the tidal habitable zone. For these planets the tidal heating flux is higher
for the masses of  model A than for  model B. This is due to the fact that in the second 
case the eccentricities of the planets are lower.  On the other hand the heating of  planet d,
which is in the optimistic habitable  zone  but not the  conservative habitable zone,
is larger for model B.  When $Q'$ is increased to $10^3$ only planet e  of model A remains  in
both the tidal and conservative habitable zones, whereas planet d remains in both the
tidal and optimistic habitable zones.

Our calculations described in subsection \ref{Numressec} indicate
convincingly that if the system is to remain near the present state
over $Gy$ time scales, $Q'$ should be in the range  $10^{2-3}$.
As we have already mentioned, the age of the TRAPPIST-1 system is very 
uncertain. 
 Adopting an age  of $5Gy,$ 
 in the middle  of the age range estimate given by   Luger at al. (2017), 
our simulations indicate that if the system has remained
as now  for that time, $Q' > \sim   10^3.$ 
For  $Q'=10^3$ 
 TRAPPIST-1 e is  in  both the
tidal and conservative  habitable zones  when the
masses  are taken to be those of  model A.
In addition we remark that TRAPPIST-1d is in the tidal and optimistic habitable zones
for $Q'< 2\times 10^{3}$ allowing ages up to $\sim 10Gy$ in that case.
However, if we adopt the lower bound of $0.5Gy$ 
for the age given by Filippazzo et al. (2015) then
we have $Q' > \sim 10^2.$ If $Q' =10^2,$ for both model A
and model B  TRAPPIST-1 e and f are in both the tidal
and conservative habitable zones as indicated above.

For the current configuration
the evolution rate 
is dominated by the energy dissipation 
rate in TRAPPIST -1 b
which exceeds that in planets  TRAPPIST -1 e  and f by two orders of magnitude  or more 
for the same $Q'.$
it is thus possible to envisage an increased 
heating rate for   TRAPPIST - 1 e   and  f    
by reducing $Q'$ for them  as compared to TRAPPIST-1 b 
by a modest factor, $f_{Q'}$ say,
that is large enough to bring  them  into the tidal 
habitable and conservative habitable zones when the system has an  age up
to $f_{Q'}$ times greater than if the reduction was not made
or $f_{Q'}\times 0.5Gy$ for the   lower bound considered in the last
paragraph. 
 We  recall  that the orbital evolution  is driven by the total energy dissipation rate
(see Section \ref{SLap}
equation (\ref{FinalLap})).

\section{Discussion and conclusion}\label{sec:discussion}

In this paper we have studied the dynamical evolution of models representing planetary systems with initial conditions
close to those inferred for the TRAPPIST-1 system.  Tidal interaction with the central star that causes orbital
circularization  has been included. This leads to the system attaining a state where  all consecutive 
groups of three planets, apart from c, d, and e,  enter into three body Laplace resonances, each of which is associated with a pair
of first order resonances for which  the  related resonant angles librate.  
The   period ratios for all consecutive pairs of planets  then  secularly increase with time. 
  
  This behaviour occurs naturally  in a system in which energy dissipation occurs while its total angular 
  momentum is conserved.  To illustrate  this process  we  provided an analytical description of  the evolution of  such a system 
  in which all consecutive triples are linked through Laplace resonances. In that case the rate of evolution 
 is determined entirely by the total rate of energy dissipation, here associated with tidal circularization,
 while period ratios increase with time. This analysis is later extended to the case where an intermediate triple
 is not connected by a Laplace resonance. In this case the system tends to evolve as two separate subsystems
 and it is necessary to specify how they are linked.
  
We  performed numerical simulations of two model systems, A,  and B,  
of the present TRAPPIST-1 system, for which different estimates for  masses have been provided.
The tidal interaction  is characterised by the tidal parameter $Q'$  associated with the planets
 for which values in the range
$Q'_0-100Q'_0$ with $Q'_0 = 0.122127$  have been considered. We considered cases for which $Q'$ was the same for all planets
and for which tidal dissipation was allowed to occur only in the innermost two planets.
These effectively yield the same results which is a consequence of the dependence of tidal dissipation 
on semi-major axis resulting in this being very much more effective for those planets.
Values of $Q'$ in the range $Q'_0-100Q'_0$ were considered with results being extrapolated for larger values.
Simulations starting from initial conditions close to the observed system  behave as indicated above.
The rate of increase of the period ratios suggests that,  in order that these do not depart significantly
from current values,   $Q'  > \sim  1.2 \times 10^3 (t_*/5 Gy),$   where $t_*$ is the age of the system.

We remark that although planets b and c as well as c and d are close to higher order resonances,
these do not appear to play a very significant role  in the present configuration
 on account of their small eccentricities.
This   was also exemplified in simulations for which the inner three planets were separated from the others
and moved slightly inwards. The inner subsystem expanded to merge with the outer planetary subsystem.
In doing so it evolved as a triple system in a Laplace resonance associated with two first order resonances
precisely as described in the analytic model. 
 When the  subsystems  merge a system  
 like TRAPPIST-1 could be produced under appropriate conditions.

In order to relate the current configuration of TRAPPIST-1 to its formation, 
we performed simulations of
the six innermost  planets migrating in a variety of protoplanetary disc
models  in Section \ref{disceffect} as this is conducive to forming resonant chains.
In one set of simulations the planets were allowed to move interior to an inner disc edge
before terminating migration and in a second set they were allowed to stall just beyond
the inner edge without entering the inner cavity.

 For reasonable  migration prescriptions,  in the first case it was found that  
 the inner two planets accelerated  away from the rest such that the present system could not be formed.
This  implies  that in that case  these  must have migrated
 inwards separately from the rest so supporting the formation of two subsystems as mentioned above.
When  it was assumed that the system  stalled just beyond the disc inner edge, noting that, 
in order to  produce a migration timescale smaller than
$ 10^6 y$  at  $ a  > 0.1 au, $  the disc mass  had to exceed a few $10^{-5} M_{\odot}$ within 5 au,  it was found that  
  period ratios  increased  to excessively large values. 
This was because of  the strength of the interaction with the disc producing orbital circularization.
This situation  was unaffected by the dispersal of the disk.
 Our results  imply  that   it is unlikely that the planetary system was
trapped beyond the disc's inner edge  for a significant time before its dispersal.

We also gave a preliminary  discussion of  the effects of the tidal dissipation indicated by our simulations
on habitability in Section \ref{Habitability}.
It is important to emphasise that this is  based on uncertain parameters and a restricted 
set of simulations that made specific assumptions about the applicable  tidal parameter $Q'$ 
for these planets that should be  relaxed in future work.
 Nonetheless our results  indicate that planet   e  
is  potentially  in  both the  conservative and  tidal habitability zones,  the latter  defined to be the regime where
plate tectonics are expected to be active, while planet d  is  in the tidal and  optimistic habitable zones. 
If the lower bound for the age of the system given by  $t_* = 0.5Gy$ holds   (Filippazzo et al. 2015),
this would also be the case for planets e and f.
In general a younger age for the system allows smaller $Q'$ associated with the planets
which accordingly makes tidal habitability more likely for the outer planets.
 
We remark that the TRAPPIST-1 planets will be accessible to atmospheric characterisation
with the JWST (James Web Space Telescope) and ELT (Extremely Large Telescope)
over the next decade. This will be a unique 
opportunity to examine the conditions present on these planets and
address issues concerning  their habitability.

\section*{Acknowledgements}
 JP and ES acknowledge support from the Polish National Science
 Centre MAESTRO grant DEC-2012/06/A/ST9/00276. 
 In addition we thank an anonymous referee for
comments that resulted in significant improvements to the original manuscript.

\begin{appendix}
\section{Expansion of systems composed of two subsystems  with Laplace resonances  }

We consider a system of $N$ planets for which the inner set for which $i = 1,2...N_1$ form a system of planets linked by three body Laplace resonances
and the outer set for which $i= N_1+1 , N_1+2,....N$ also form a set linked by three body Laplace resonances.
In this case planets $i = N_1$ and $i= N_1+1$ can still be   resonant  but linked by only one, rather than two first order resonances, precluding Laplace resonances
of the pair with  both  planets $i = N_1+2$ and $N_1-1.$  

We begin by first considering the inner system for which equations (\ref{nomenc2a}) and (\ref{nomenc3a})  apply with $N$ replaced by $N_1.
$These equations yield

\begin{align}
& \frac{d E_a}{dt }  - \alpha_a \frac{d E_{1}}{dt}= \sqrt{GM}
\frac{d\Delta_{1}}{dt}\sum_{k=0}^{N_1-2}\left( \frac{m_{k+2}}{3}\sqrt{a_{k+2}} \sum_{j=0}^k \Theta_{1,j}\right) \label{nomenc41}
\end{align}

\begin{align}
& {\rm and}\hspace{3mm} \frac{d J_a}{dt }  - \beta_a \frac{d J_{1}}{dt}
 = \frac{d\Delta_{1}}{dt}  \sum_{k=0}^{N_1-2}\left(  \frac{m_{k+2}}{3}a_{k+2}^2
\sum_{j=0}^k \Theta_{1,j}\right).\label{nomenc51}
\end{align}
where the total energy, $E$  and angular momentum, $J$  of the  inner subsystem have been  taken to be  $E_a$ and $J_a$ respectively. In addition
$\alpha$ and $\beta$  have been  replaced with


\begin{equation}
 \hspace{-0cm}\alpha_a = \sum_{k=-1}^{N_1-2}\left(\frac{m_{k+2}}{m_{1}} \right) \sqrt{\frac{ a_{k+2}} {a_{1}}} \hspace{2mm} {\rm and}
\end{equation}
 
\begin{equation}
\hspace{-0cm}\beta_a =  \sum_{k=-1}^{N_1-2}\left(\frac{m_{k+2}}{m_{1}} \right) \left(\frac{ a_{k+2}} {a_{1}}\right)^{2}.
\end{equation}
Similarly   equations (\ref{nomenc2a}) and (\ref{nomenc3a}) can be applied to the outer planets by replacing planet, $1,$ by planet $N_1$
and ensuring that the summations are over  planets with $i$ exceeding   $N_1,$ with the result that

\begin{align}
& \frac{d E_b}{dt }  - \alpha_b \frac{d E_{N_1}}{dt}=\nonumber\\ 
&\sqrt{GM}
\frac{d\Delta_{N_1}}{dt}\sum_{k=0}^{N-N_1-1}\left( \frac{m_{N_1+k+1}}{3}\sqrt{a_{N_1+k+1}} \sum_{j=0}^k \Theta_{N_1,j}\right), \label{nomenc61}
\end{align}

\noindent and

\begin{align}
& \frac{d J_b}{dt }  - \beta_b \frac{d J_{N_1}}{dt}
 =\nonumber\\ 
&\frac{d\Delta_{N_1}}{dt}  \sum_{k=0}^{N-N_1-1}\left(  \frac{m_{N_1+k+1}}{3}a_{N_1+k+1}^2
\sum_{j=0}^k \Theta_{N_1,j}\right).\label{nomenc71}
\end{align}
where the total energy, $E$  and angular momentum, $J$  of the outer system have been  taken to be  $E_b$ and $J_b$ respectively.
But  it is important to note that this  outer system contains planet $N_1$ in addition to the outer subsystem. In addition
in this case $\alpha$ and $\beta$  have been  respectively replaced with

\begin{equation}
 \hspace{-0cm}\alpha_b = \sum_{k=-1}^{N-N_1-1}\left(\frac{m_{N_1+k+1}}{m_{N_1}} \right) \sqrt{\frac{ a_{N_1+k+1}} {a_{N_1}}}\hspace{2mm} {\rm and}
\end{equation}

\begin{equation}
\hspace{-0cm}\beta_b =  \sum_{k=-1}^{N-N_1-1}\left(\frac{m_{N_1+k+1}}{m_{N_1}} \right) \left(\frac{ a_{N_1+ k+1}} {a_{N_1}}\right)^{2}.
\end{equation}

Adding (\ref{nomenc41}) and (\ref{nomenc61})  we obtain
\begin{align}
& \frac{d E}{dt }  - \alpha_a \frac{d E_{1}}{dt}  - (\alpha_b-1) \frac{d E_{N_1}}{dt}=\nonumber\\
&\sqrt{GM}
\frac{d\Delta_{1}}{dt}\sum_{k=0}^{N_1-2}\left( \frac{m_{k+2}}{3}\sqrt{a_{k+2}} \sum_{j=0}^k \Theta_{1,j}\right) \nonumber \\
&+ \sqrt{GM}
\frac{d\Delta_{N_1}}{dt}\sum_{k=0}^{N-N_1-1}\left( \frac{m_{N_1+k+1}}{3}\sqrt{a_{N_1+k+1}} \sum_{j=0}^k \Theta_{N_1,j}\right), \label{nomenc81}
\end{align}
where the total energy of   the whole system is now, $ E= E_a+E_b -E_{N_1}.$ 
Similarly, 
adding (\ref{nomenc51}) and (\ref{nomenc71})  we obtain

\begin{align}
& \frac{d J}{dt }  - \beta_a \frac{d J_{1}}{dt} - (\beta_b-1) \frac{d J_{N_1}}{dt}
 =\nonumber\\
&\frac{d\Delta_{1}}{dt}  \sum_{k=0}^{N_1-2}\left(  \frac{m_{k+2}}{3}a_{k+2}^2
\sum_{j=0}^k \Theta_{1,j}\right)\nonumber \\ 
 &+\frac{d\Delta_{N_1}}{dt}  \sum_{k=0}^{N-N_1-1}\left(  \frac{m_{N_1+k+1}}{3}a_{N_1+k+1}^2
\sum_{j=0}^k \Theta_{N_1,j}\right),\label{nomenc911}
\end{align}
where the total angular momentum of  the whole system is  $ J= J_a+J_b -J_{N_1}.$

We now connect $dE_1/dt$ and $dE_{N_1}/dt $ by taking the time derivative of 
(\ref{nomenca}) with $i=1$ and $k=N_1-2$ which  then takes the form
\begin{equation}
n_{1}-n_{N_1}= \Delta_{1} \sum_{j=0}^{N_1-2}\Theta_{1,j}.   \label{nnnnnn}
\end{equation}
  from this   we find that 
\begin{align}
&  \alpha_a \frac{d E_{1}}{dt}  +( \alpha_b-1) \frac{d E_{N_1}}{dt}=\nonumber\\
& -\gamma\frac{dn_1}{dt}+\frac{1}{3}\frac{d\Delta_{1}}{dt}(\alpha_b -1)m_{N_1}a_{N_1}^2n_{N_1}\sum_{j=0}^{N_1-2} \Theta_{1,j},
\label{nomenc811}
\end{align}
where
\begin{align}
&  \gamma= \frac{1}{3}\left(\alpha_a m_{1}a_{1}^2n_{1}+(\alpha_b-1) m_{N_1}a_{N_1}^2n_{N_1}\right)\equiv \frac{1}{3}m_1n_1a_1^2\alpha
\label{nomenc101}
\end{align}
Recalling that we are working in the limit of very small eccentricities 
the time derivative of (\ref{nomenca})
can be  expressed in terms of the time derivatives of the  angular momenta, $J_1$ and $J_{N_1}$  
 in which case we obtain

\begin{align}
&  \beta_a \frac{d J_{1}}{dt}  + (\beta_b-1) \frac{d J_{N_1}}{dt}\nonumber\\
&= -\delta \frac{dn_1}{dt}+\frac{1}{3}\frac{d\Delta_{1}}{dt}(\beta_b-1) m_{N_1}a_{N_1}^2\sum_{j=0}^{N_1-2} \Theta_{1,j},
\label{nomenc111}
\end{align}
where
\begin{align}
&  \delta= \frac{1}{3}\left(\beta_a m_{1}a_{1}^2+(\beta_b -1)m_{N_1}a_{N_1}^2\right)\equiv \frac{1}{3}m_1a_1^2\beta,
\label{nomenc121}
\end{align}
and we remark that 
 $\gamma/\delta = n_1\alpha/\beta.$

We now use (\ref{nomenc811}) and (\ref{nomenc111}) to eliminate $dE_1/dt$ together with  $dE_{N_1}/dt$ from (\ref{nomenc81}),  and
  $ dJ_1/dt$ together with  $dJ_{N_1}/dt$ from 
 (\ref{nomenc911}) respectively.
Finally we eliminate $dn_1/dt$ from the resulting 
pair of equations to obtain a single equation relating the rates of change of  $\Delta_1$ and $\Delta_{N_1}$ to the total rate of energy dissipation.
After some algebra this can be shown to take the form 
\begin{align}
&\frac{d E}{dt}  -\frac{n_1\alpha}{\beta}\frac{d J}{dt} =\nonumber\\  
& \frac{d\Delta_{1}}{dt}\left.\sum_{k=0}^{N_1-2} \frac{m_{k+2}a_{k+2}^2}{3}  \left(  n_{k+2} -\frac{n_1\alpha}{\beta}\right)
\sum_{j=0}^k \Theta_{1,j}\right.+\nonumber\\
&\frac{d\Delta_{N_1}}{dt}\left.  \sum_{k=0}^{N-N_1-1}\frac{m_{N_1+k+1}a_{N_1+k+1}^2}{3}\left( n_{N_1+k+1}  
-\frac{n_1\alpha}{\beta}\right)
\sum_{j=0}^k \Theta_{N_1,j}+\right.\nonumber\\
&\frac{d\Delta_{1}}{dt} \sum_{j=0}^{N_1-2} \Theta_{1,j} \left.
 \sum_{k=0}^{N-N_1-1}\frac{m_{N_1+k+1}a_{N_1+k+1}^2}{3}\left( n_{N_1+k+1}   - \frac{n_1\alpha}{\beta}  \right)\right..
\label{Edotdouble} \end{align}
Note that when $N_1=1$ we recover the case of a single system. In that case 
only the term  proportional 
to $d\Delta_{N_1}/dt$ contributes such that equation (\ref{Edotsingle}) is recovered.
\subsection{ Separately expanding subsystems}
Equation (\ref{Edotsingle}) 
 is also recovered if $\Delta_1/\Delta_{N_1}$ is assumed to be  constant corresponding to the entire system expanding  with the ratio of the difference 
 between the mean motions of consecutive pairs remaining constant, but not necessarily the same constant,
with the proviso that we use
\begin{equation}
\Theta_{i,k} =  \frac{\Delta_{i+k}}{\Delta_{i}} 
 = \frac{n_{i+k}-n_{i+k+1}}{n_{i}-n_{i+1}}
 \hspace{3mm} {\rm for} \hspace{3mm} k=1,2... N-i-1,
\end {equation}
but  {\it do not use}  the second equality in (\ref{nomenc}) which only holds for Laplace resonances.

Proceeding in this way in general, without assuming $\Delta_{N_1}/\Delta_1$ is constant,  (\ref{Edotdouble})  can be rewritten in the alternative form
\begin{align}
&\frac{d E}{dt}  -\frac{n_1\alpha}{\beta}\frac{d J}{dt} =   \frac{d\Delta_{1}}{dt}\left(\sum_{k=0}^{N-2} \frac{m_{k+2}a_{k+2}^2}{3}  \left(  n_{k+2} -\frac{n_1\alpha}{\beta}\right)
\Lambda_k\right)
\label{Edotdouble1}
 \end{align}
where
\begin{align}
&\Lambda_k= \frac{ n_1-n_{k+2}}{n_1-n_2} \hspace{3mm}{\rm for} \hspace{3mm} k  <  N_1 -1,\hspace{3mm} {\rm and}
\end{align}
\begin{align}
&\Lambda_k=  \frac{n_{N_1}-n_{k+2}}{n_{N_{1}} --n_{N_1+1}}\frac{d\Delta_{N_1}}{d\Delta_1} + \frac{n_1 - n_{N_1}}{n_{1}-n_{2}}\hspace{3mm}{\rm for} \hspace{3mm} k\ge N_1-1.
\end{align}
We remark that when $\Delta_{N_1}$ and $\Delta_1$ increase monotonically with time $\Lambda_k $ is easily shown to be
positive definite and an increasing function of $k.$
This is sufficient to ensure that the quantity multiplying $d\Delta_1/dt$ in (\ref{Edotdouble1}) is negative.
It then follows that if the total angular momentum of the system is conserved, energy dissipation drives its progressive expansion.
\end{appendix}

\end{document}